%% file: p0-main.tex
\documentclass{EG_files/egpubl}
 \usepackage{silence}
\JournalSubmission    
\usepackage[T1]{fontenc}
\usepackage{EG_files/dfadobe}  
\usepackage{amsfonts}  
\usepackage{amsmath}

\usepackage{cite}  
\BibtexOrBiblatex
\electronicVersion
\PrintedOrElectronic
\ifpdf \usepackage[pdftex]{graphicx} \pdfcompresslevel=9
\else \usepackage[dvips]{graphicx} \fi

\usepackage{EG_files/egweblnk} 

\usepackage{mwe}
\usepackage{subfig}
\usepackage{xcolor}
\usepackage{tikz}
\usepackage{pgfplots}

\title[Evocube: a Genetic Labeling Framework for Polycube-Maps]%
      {Evocube: a Genetic Labeling Framework for Polycube-Maps}

\author[C. Dumery et al.]
{\parbox{\textwidth}{
        \centering C. Dumery$^{1, 4}$\orcid{0000-0001-5314-7979}, F. Protais$^2$\orcid{0000-0002-2089-3745}, S. Mestrallet$^1$\orcid{0000-0002-4519-2814}, C. Bourcier$^1$\orcid{0000-0001-6171-024X}, F. Ledoux$^3$\orcid{0000-0003-3469-3186}
        }
        \\
{\parbox{\textwidth}{\centering $^1$CEA, Université Paris-Saclay\\ $^2$Université de Lorraine, CNRS, Inria, LORIA \\ $^3$CEA DAM, LIHPC, Université Paris-Saclay \\ $^4$EPFL, School of Computer and Communication Sciences
       }
}
}

\pgfplotsset{compat=1.17}
\begin{document}

\teaser{
 \includegraphics[width=1.0\linewidth]{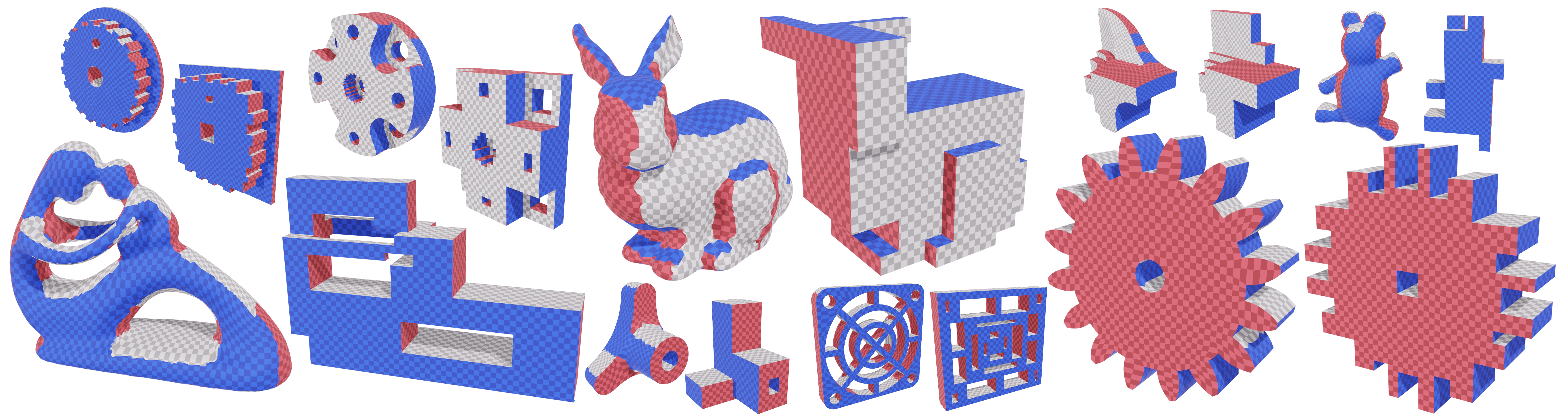}
 \centering
  \caption{Collection of labelings and polycube-maps of general shapes generated using the Evocube framework.}
\label{fig:teaser}
}

\maketitle
\input{p1-abstract.tex}

\input{p2-introduction.tex}
\input{p3-related-work.tex}

\input{p4-genetic-labeling.tex}

\input{p5-polycube-maps.tex}
\input{p6-experiments.tex}

\input{p7-conclusion.tex}

\bibliographystyle{EG_files/eg-alpha-doi}  
\bibliography{evocube.bib}        


\input{p8-appendix.tex}


\end{document}

%% file: p1-abstract.tex
\begin{abstract}
   
Polycube-maps are used as 
base-complexes in various fields of computational geometry, including the generation of regular all-hexahedral meshes free of internal singularities.
However, the strict alignment constraints behind polycube-based methods make their computation challenging for 
CAD models used in numerical simulation via Finite Element Method (FEM). 
We propose a novel approach based on an evolutionary algorithm to robustly compute polycube-maps in this context. \\
%
We address the labeling problem, which aims to precompute polycube alignment by assigning
one of the 
base axes to each boundary face on the input.
Previous research has described ways to initialize and improve a labeling via greedy local fixes. However, such algorithms lack robustness and often converge to inaccurate solutions for complex geometries.
Our proposed framework alleviates this issue by embedding labeling operations in an evolutionary heuristic,
defining fitness, crossover, and mutations in the context of labeling optimization.
%
We evaluate our method on a thousand smooth and CAD meshes, showing Evocube converges to accurate labelings on a wide range of shapes. The limitations of our method are also discussed thoroughly.  

\ccsdesc[500]{Mathematics of computing~Mesh generation}
\ccsdesc[500]{Mathematics of computing~Evolutionary algorithms}



\printccsdesc   
\end{abstract}  

%% file: p2-introduction.tex
\section{Introduction}

\begin{figure}[tbp]
  \centering
  \includegraphics[width=0.9\linewidth]{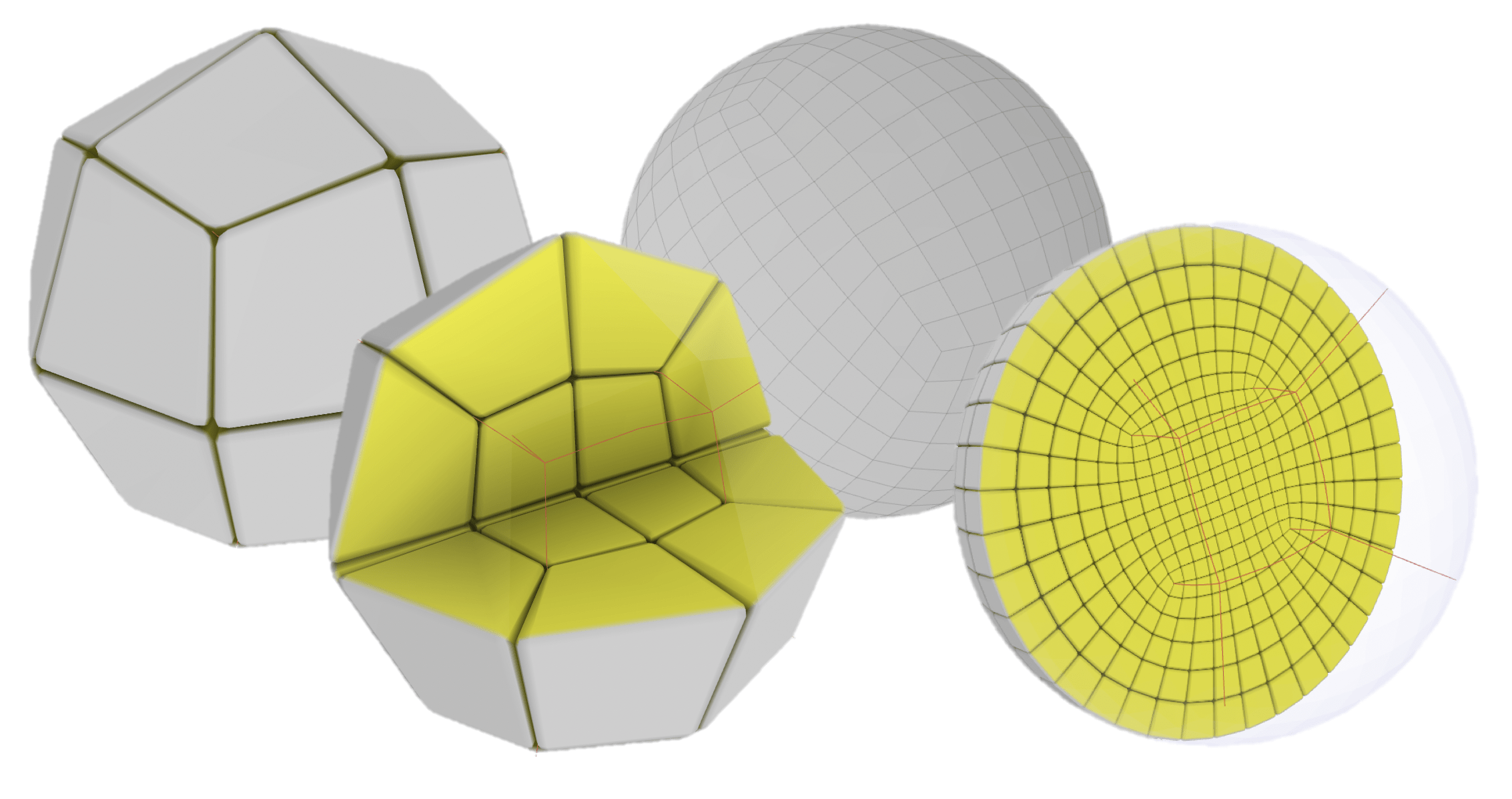}
  \caption{\label{fig:blocking}
          Block-structured hexahedral mesh of a sphere, structure on the left and resulting mesh on the right.}
\end{figure}

Scientific computational analysis based on finite element method (FEM) or finite volume method (FVM) is increasingly used to model engineering problems. Recent developments incorporate coupled physical phenomena and require complex geometric shapes. However, FEM and FVM are limited by volumetric mesh generation. In practice, for many cases of interest, all-hexahedral meshes, or hex meshes, are preferred over tetrahedral meshes \cite{Shepherd2008}. Their use drastically reduces computational cost and memory footprint while preserving numerical accuracy. Although tetrahedral meshing is now robustly performed on general 3D shapes, direct generation of good-quality hex meshes remains an open problem. To ensure high-accuracy and convergence speed, finite element analysis requires block-structured hex meshes with low cell distortion, as illustrated in Figure \ref{fig:blocking}. As of today, such properties cannot be guaranteed for general shapes.

All-hexahedral meshing has been studied for several decades. Several approaches have emerged, relying on different techniques: advancing front\cite{plastering_1,plastering_2,hmorph,tautges95whisker,ledoux07}, overlay-grid or octree-based \cite{schneiders1996refining,Marechal2009,zhang_jessica10,gao2019feature}, frame field\cite{huang2011boundary,li_liu12, ray2016practical, kowalski2016smoothness,palmer2019algebraic}, medial axis\cite{sheffer1999hexahedral, LMPS16, Quadros}, or polycube-maps
\cite{tarini2004polycube,gregson_polycube_2011,polycut_livesu}. As of today, none of these methods has been successful in automatically generating satisfying hex meshes for realistic 3D shapes. 
In practice, time-consuming domain partitioning is often performed interactively \cite{parrish2007,takayama2019, calderan2019, LiInteractivePolycube} instead. For the generation of block-structured meshes, we believe frame field and polycube-map methods are the most promising directions in state-of-the-art research. Frame fields are generated by optimizing a smoothness energy 
and yield the desired block-structure for simple geometries, but fail on more complex shapes. Polycube-maps, on the other hand, succeed on a broader spectrum of 3D shapes but do not provide a usable block structure since interior singularities are missing.
Polycube-map generation is performed by deforming a 3D shape to obtain an orthogonal polyhedron (or \emph{polycube}~\cite{tarini2004polycube}), and generating a volumetric map $p$ between them.
A hex mesh is then extracted from the polycube-map via subdivision following the integer grid, and morphed back onto the initial 3D shape following the inverse map $p^{-1}$. By construction, resulting meshes lack interior singularities and low-quality hexes are generated close to the boundary. In state-of-the-art pipelines, this issue is partially addressed with post-processing improvements that insert layers of cells along the boundary \cite{kowalski2012fun,cherchi2019selective}.

\begin{figure}
\begin{minipage}{.33\linewidth}
\centering
\subfloat[compact \\ $N_c = 8, D_a = 1.100$]{\label{fig_a}\includegraphics[width=\linewidth]{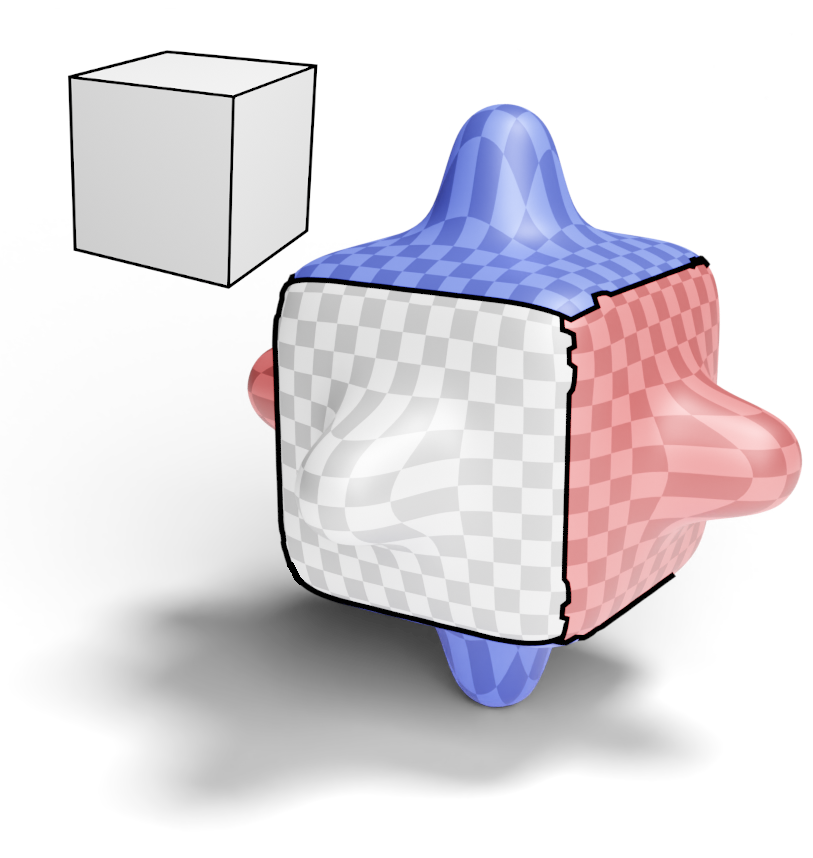}}
\end{minipage}%
\begin{minipage}{.33\linewidth}
\centering
\subfloat[high fidelity \\ $N_c = 56, D_a = 1.047$]{\label{fig_b}\includegraphics[width=\linewidth]{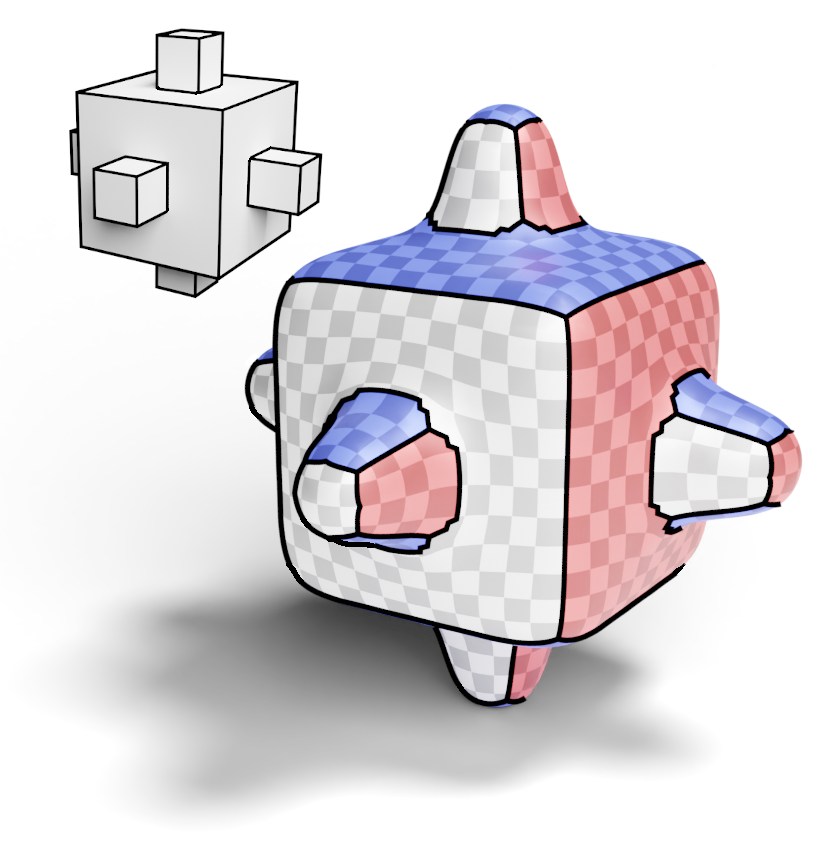}}
\end{minipage}
\begin{minipage}{.33\linewidth}
    \centering
    \subfloat[non-monotone\\ $N_c = 56, D_a = 1.723$]{\label{fig_c}\includegraphics[width=\linewidth]{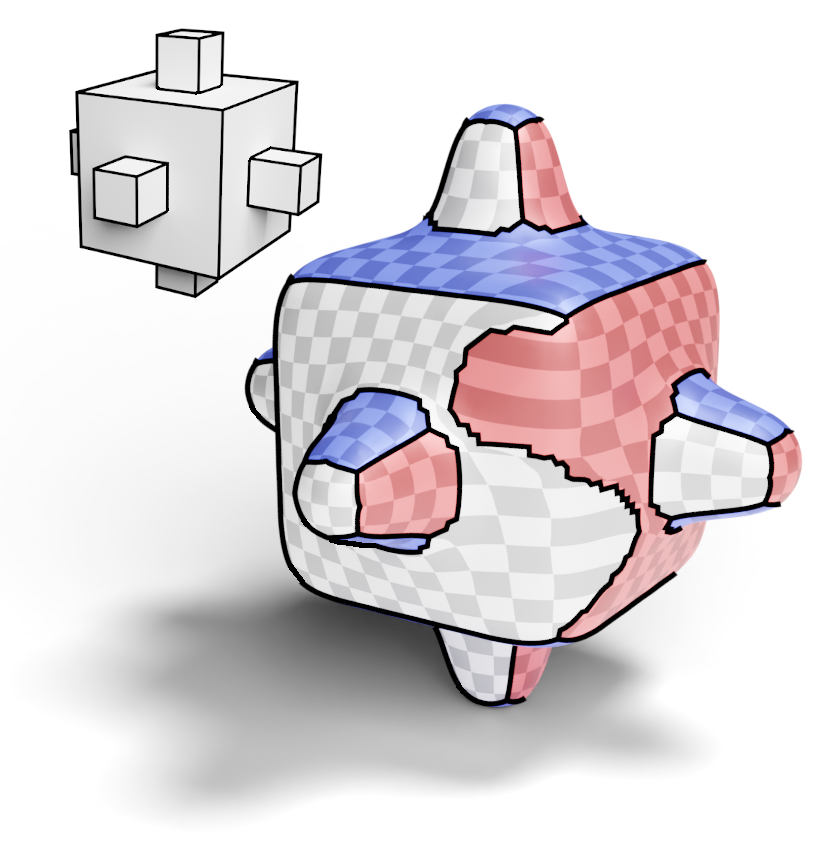}}
\end{minipage}%
\caption{Illustration of labeling properties and associated polycubes. We report the number of corners in the polycube $N_c$ and the area distortion $D_a$ \cite{tarini2004polycube}.}
\label{fig:polycut_terminology}
\end{figure}

We observe that polycube alignment can be quickly estimated and evaluated, and propose a novel approach based on an evolutionary algorithm \cite{book_evolutionay} to robustly compute \textit{polycube labelings}. 
Starting from a tetrahedral mesh $T_\Omega$ of the 3D domain $\Omega$, polycube labeling consists in assigning one of the six base axis directions $\{\pm X,\pm Y, \pm Z\}$ to each face of the boundary $\partial T_\Omega$. Adjacent triangles that are assigned the same direction form a \textbf{chart}. 
Following the terminology introduced by Livesu et al. \cite{polycut_livesu}, a labeling should define a \textbf{valid} polycube topology\cite{eppstein2010steinitz, sokolov2015fixing},
and compromise between the following quality criteria illustrated in Fig. \ref{fig:polycut_terminology}:
\begin{itemize}
    \item \textbf{Fidelity}: angles between assigned directions and triangle normals remain low;
    \item \textbf{Compactness}: the number of charts is small;
    \item \textbf{Monotonicity}: boundaries between charts are exempt from \textit{turning points}, i.e. significant changes in boundary direction.
\end{itemize}

\noindent

The nature of the polycube labeling problem makes genetic methods a good fit. In previous work, labelings were computed using deterministic operations which explored a limited set of solutions. We observe that generating and evaluating large quantities of labelings is significantly less costly than computing low-distortion polycube-maps for all possibilities. We thus propose an evolutionary framework that explores a wider range of solutions than previous methods.
Starting from an initial solution, and over several iterations or \textit{generations}, we generate new solutions referred to as \textit{individuals}. At any given time, the current set of individuals being considered is referred to as \textit{population} and is stored in a fixed-size \textit{archive}. We measure the quality of an individual with a \textit{fitness} function. Each generation, some individuals are sampled from the archive to undergo labeling modifications named \textit{mutations}. Individuals with high fitness are then selected for \textit{crossover}, where a pair of individuals generates a third one combining both of its parents traits. At the end of each generation, the fittest individuals are inserted back into the archive. This process aims to improve population quality over generations. If the content of the archive does not evolve for several consecutive generations, then we consider that the algorithm has converged and stop the process. For a large majority of inputs, \textit{Evocube} converges towards a labeling associated with a low-distortion polycube within a dozen generations.

We validate our algorithm on a large collection of over one thousand CAD and smooth models, showcasing some in Figure \ref{fig:teaser}. \textit{Evocube} produced labelings leading to low-distortion polycubes on an overwhelming majority of inputs, as reported in Section \ref{sec:experiments}. We also analyze failure cases and comment on the limitations of our method.


%% file: p3-related-work.tex

\subsection{Related Work}

\begin{figure*}[htbp]
  \centering
  \mbox{} \hfill
  \includegraphics[width=1.0\linewidth]{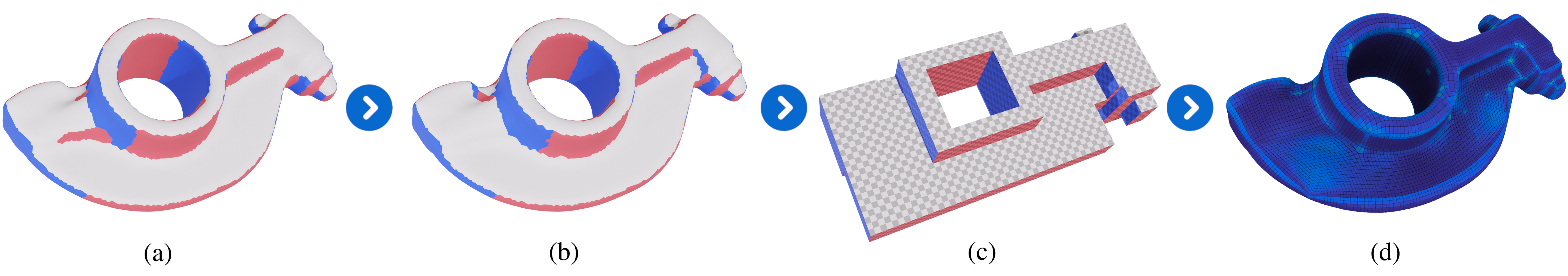}
  \caption{\label{fig:pipelineOverview}
    Starting from a triangular mesh representation of a 3D shape, (a) we generate an initial non-optimal and potentially invalid labeling using a graph-cut method\cite{polycut_livesu}; (b) this labeling is then refined within our Evocube framework; (c) we generate a polycube with topology matching our labeling, and (d) a grid-like mesh is extracted, padded and morphed back onto the initial 3D shape.}
\end{figure*}

\textbf{Polycube labeling.}
Polycubes were first used in computer graphics for seamless texturing of triangulated surfaces \cite{tarini2004polycube}. They rely on a polyhedral structure and a volumetric map, which can either be computed one after the other, 
or together via mesh deformation towards the orthogonal polyhedron closest to the input 3D shape.
The polyhedral structure is usually defined by labeling  each element of  $\partial T_\Omega$ with a value that represents one of the six base axes $\{\pm X, \pm Y, \pm Z\}$. A naive labeling can be computed by assigning to each surface triangle the label closest to its normal ~\cite{gregson_polycube_2011}. However, this does not produce a valid structure in general and additional labeling refinement is necessary. This refinement is guided by a set of sufficient topological conditions for the existence of a valid polyhedron defined by Eppstein et al. \cite{eppstein2010steinitz} and used by Livesu et al. \cite{polycut_livesu}. Similarly, Hu et al. \cite{cvt_pc_zhang2016} use a modified Centroidal Voronoï Tesselation labeling in the space of normals, followed by a post-processing stage to sanitize topological inconsistencies.

\textbf{Polycube deformation.}
In previous work, mesh deformation is used to compute an approximate polycube by minimizing an energy that penalizes surface normals poorly aligned with base axes, until the polycube structure is revealed \cite{pc_opt_cad2014,l1pc2014,closedform_pc2016}. Some methods also start from an initial labeling ~\cite{gregson_polycube_2011, polycut_livesu} to assist the deformation, or interleave both approaches by updating a target labeling between each deformation iteration \cite{fu2016efficient}. 
Mesh deformation may lead to strongly distorted or flipped elements \cite{gregson_polycube_2011,l1pc2014}. In recent work \cite{guo2020cut}, the AMIPS \cite{amips} energy guarantees an inversion-free deformation by diverging in the presence of inverted cells.
However, in some degenerate cases and due to inconsistent polyhedron topology, additional post-processing stages are necessary. Unfortunately, Sokolov et al. \shortcite{sokolov2015fixing} show that degenerate conditions are not merely local and the computation of a globally valid structure remains a challenging problem.

\textbf{Hex quality improvements.} Hex meshes generated via polycube-maps do not have inner singularities. As a consequence, state-of-the-art polycube-based pipelines insert one or more layers of hexes along the whole boundary~\cite{gregson_polycube_2011}. Such a process gives more degrees of freedom to smooth the mesh by pushing singularities inside. Unfortunately, naive global insertion can also locally decrease mesh quality. Kowalski et al. \cite{kowalski2012fun} define three types of fundamental layers that can be added locally to better capture boundary curves and surfaces solving an integer linear program. Similarly, Cherchi et al. \cite{cherchi2019selective} define selective padding which is able to add hex layers with a quality improvement guarantee. Recently, Guo et al. \cite{guo2020cut} enhance polycube-maps through clever cuts directly on the polycube, achieving similar results. Authors of~\cite{robust_remesh2017} propose an approach to simplify a mesh using its base complex structure. While this approach is more dedicated to optimize a mesh generated by an overlay-grid approach, it could be used to post process Polycube-like meshes in order to extract coarse structure.

\textbf{Machine learning and mesh generation.} 
A variety of geometry and mesh generation problems have seen great advancements in the past decade via machine learning techniques \cite{xu2012,panozzo_data_driven, LIM2020, dielen2021learning_direction_fields}. 
Marcias et al. ~\cite{panozzo_data_driven} describe a novel interactive method to suggest quad surface meshes to final users. 
Lim et al. \cite{LIM2020} designed an evolutionary algorithm that performs automatic blocking of a 2D manifold. They define a set of simple genetic operators on a set of points from which they robustly extract a quad layout. This layout is then evaluated and used to rank a population of sets of points. Their work achieves near-optimal blocking configurations after a large number of generations. 
Starting from an initial population of 3D models, Xu et al. \cite{xu2012} use an evolutionary algorithm to generate novel shapes. The process is interactively driven and user preferences define the fitness function. Shapes are described as an assembly of parts and the crossover operation swaps different parts between parent shapes.
Similarly, we use an evolutionary algorithm to perform polycube labeling. While the main principle and terminology are similar, the nature of the problems induces very different choices. In particular, the highly non-local nature of polycube labeling prevents the use of a naive per-triangle crossover operator. 

\subsection{Main contributions and pipeline overview}

Our contributions are embedded in a full hex meshing pipeline as illustrated in Figure~\ref{fig:pipelineOverview}.

\begin{itemize}
\item In Section \ref{sec:genetic}, we define \textit{Evocube}, an extensible evolutionary-based framework for polycube labeling; 
\item In Section \ref{sec:def_extr}, we compute polycube-maps matching a desired labeling, and derive low distortion all-hex meshes.
\end{itemize}

\noindent
Finally, in Section \ref{sec:experiments}, we evaluate our method on over a thousand natural and CAD models, and discuss its limitations. 
To foster future research on polycube labeling, our implementation of \textit{Evocube} is open-source and can easily be extended: \url{https://github.com/LIHPC-Computational-Geometry/evocube}


%% file: p4-genetic-labeling.tex

\section{Genetic labeling optimization} \label{sec:genetic}


Our method aims to generate a low-distortion polycube from an input domain. Precisely, the input to our algorithm is:
\begin{itemize}
    \item a set of three-dimensional vertices $\mathcal{V}$;
    \item a set of tetrahedral cells $\mathcal{T}_\Omega$ connecting points in $\mathcal{V}$, along with its boundary triangular mesh $\partial\mathcal{T}_\Omega$;
    \item the desired edge length $l_e$ that defines the size of a cube in our polycube-map.
\end{itemize}

We aim to compute new vertex positions $\mathcal{V}'$ 
such that all the normals in $(\mathcal{V}', \partial\mathcal{T}_\Omega)$ are aligned with one of the three main axes while $(\mathcal{V}', \mathcal{T}_\Omega)$ remains a low-distortion deformation of $(\mathcal{V}, \mathcal{T}_\Omega)$. 
For such a problem to be valid, we assume that $\mathcal{T}_\Omega$ defines a volume, is free of self-intersections, and $\partial\mathcal{T}_\Omega$ is manifold.
Specifically, our work addresses the labeling task, which consists in precomputing polycube topology directly on $\partial\mathcal{T}_\Omega$ with a labeling vector $\ell \in \{\pm X, \pm Y, \pm Z\}^{|\partial\mathcal{T}_\Omega|}$ assigning any of the six possible orientations to each boundary triangle.

Intuitively, $\ell$ subdivides $\partial\mathcal{T}_\Omega$ in a set of charts $\mathcal{C}$ comprised of neighboring triangles sharing target orientation. Charts are separated by a set of edge boundaries $\mathcal{B}$. In the event that a boundary $b \in \mathcal{B}$ is not straight, we use the method of Livesu et al. \cite{polycut_livesu} described in Section \ref{sec:init_sol} to identify turning points.


Previous work \cite{gregson_polycube_2011, polycut_livesu} has shown the feasibility of precomputing such a labeling of boundary faces using a set of modification operations that greedily improve an initial labeling. 
Our key insight is to embed local labeling fixes and associated quality criteria in a novel genetic framework. We recast the labeling problem as one of optimization with an objective function, and define genetic operations in the context of polycube labeling. Stochastic selection and crossover of candidate solutions allow \textit{Evocube} to simultaneously consider several search directions and reduce susceptibility to local minima.






The main challenge with an approach based on optimization using a heuristic is the definition of an appropriate fitness function. In the case of a genetic framework, fitness evaluation is often a time bottleneck as every individual generated needs to be ranked and assessing the quality of a solution is not straightforward. Ideally, polycube distortion with regards to the input mesh could be used as a labeling quality metric. However, computing a satisfying volumetric polycube from a labeling is time-consuming and cannot be performed repeatedly. 
Hence, we define distortion proxies and labeling modifications that guide us towards interesting solutions in the set of all possible labelings
. 

\subsection{Fitness of a labeling} \label{sec:fitness}

Evaluating labeling quality is a complex problem that encompasses several aspects of a labeling and therefore requires the definition of several metrics. In the following, we distinguish between validity conditions - which must be fulfilled for a solution to be considered feasible - and optimization criteria - which we aim to minimize.

\textbf{Validity proxy.} 
A labeling is said to be \textit{valid} if there exists a matching polycube polyhedron. Unfortunately, evaluating the validity of a labeling for a general shape remains an open problem, and the partial solutions provided in previous work \cite{eppstein2010steinitz,polycube_shape_space} are too computationally costly to be used repeatedly in a heuristic. Instead, we define a proxy validity criterion which offers no theoretical guarantee but can be efficiently evaluated.

Considering a set of charts and boundaries, the number of boundaries intersecting at a given vertex is referred to as the vertex's valency. Let us consider the following sets:
\begin{itemize}
    \item invalid corners $\mathcal{V}_{\text{inv}}$ with valency at least 4; 
    \item invalid boundaries $\mathcal{B}_\text{inv}$ between charts with opposite labels;
    \item invalid charts $\mathcal{C}_{\text{inv}}$ with strictly fewer than 4 neighbors.
\end{itemize}

\begin{figure}[htb]
\begin{minipage}{.33\linewidth}
\centering
\subfloat[$\mathcal{C}_{\text{inv}} \neq \emptyset $]{\label{fig_inv1}\includegraphics[width=\linewidth]{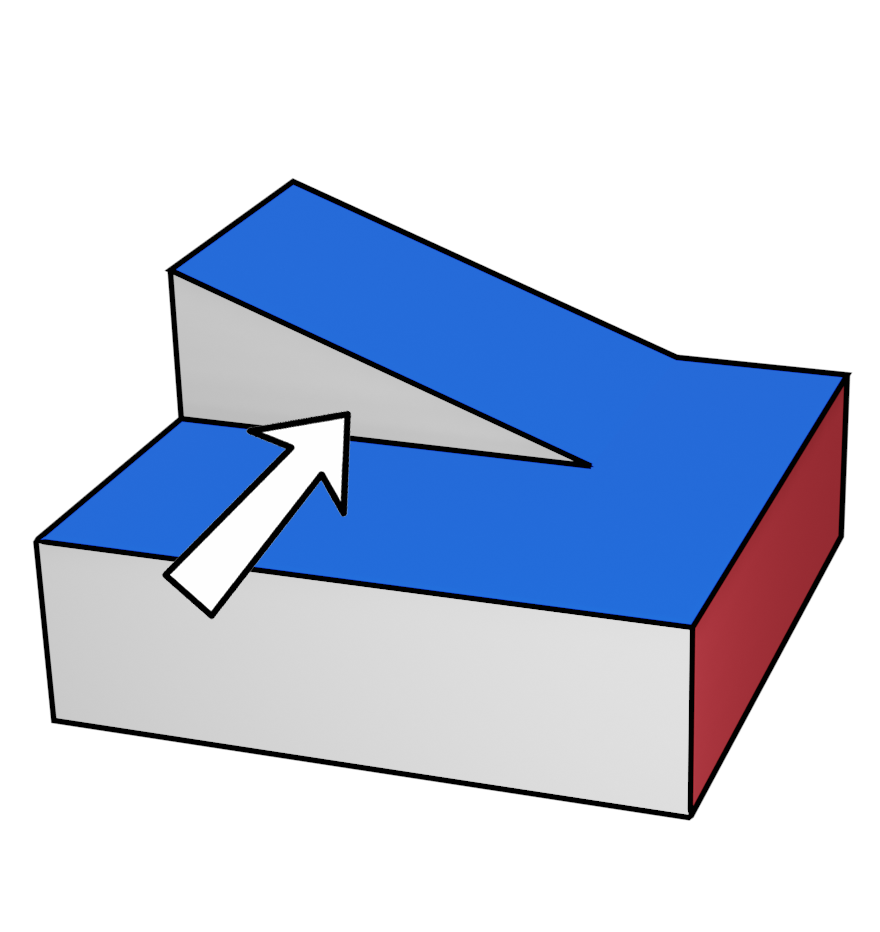}}
\end{minipage}%
\begin{minipage}{.33\linewidth}
\centering
\subfloat[$\mathcal{B}_\text{inv} \neq \emptyset $]{\label{fig_inv2}\includegraphics[width=\linewidth]{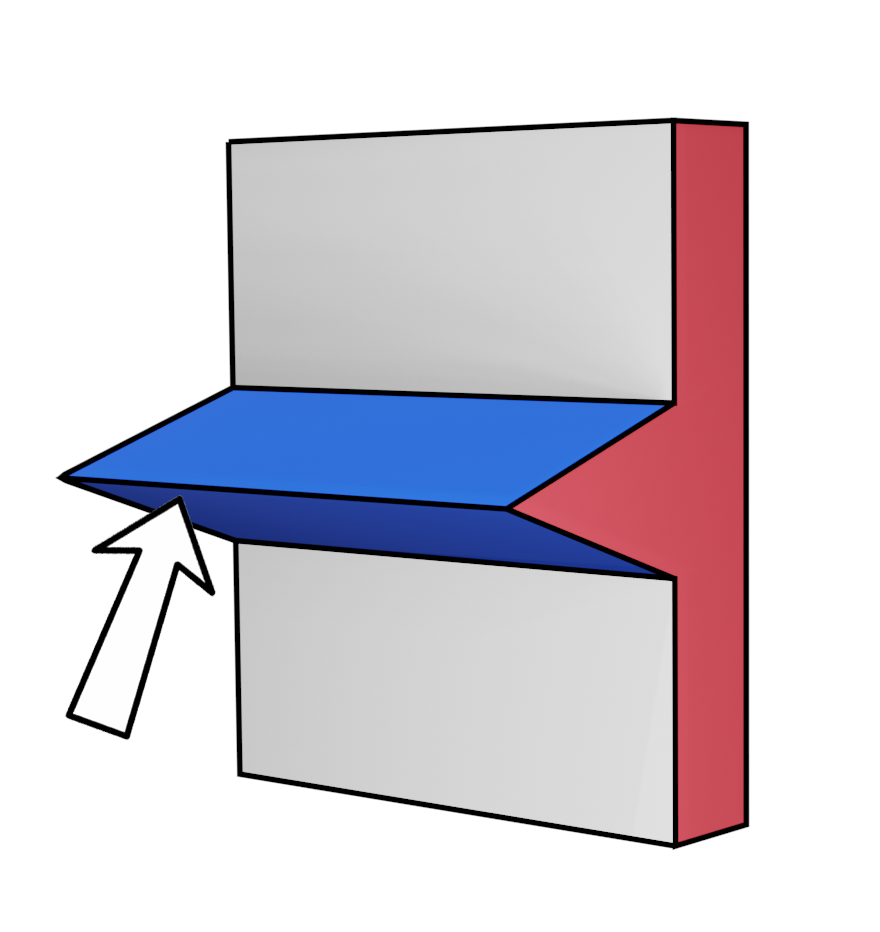}}
\end{minipage}
\begin{minipage}{.33\linewidth}
    \centering
    \subfloat[$\mathcal{V}_{\text{inv}} \neq \emptyset $]{\label{fig_inv3}\includegraphics[width=\linewidth]{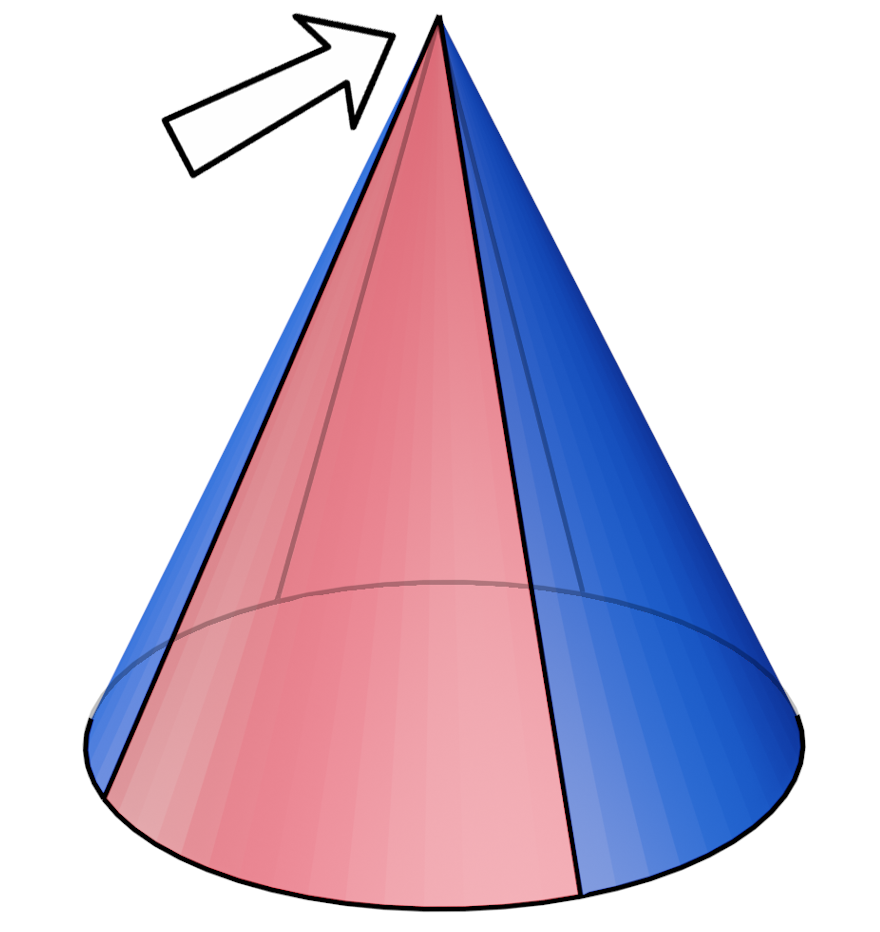}}
\end{minipage}%
\caption{Example labelings that do not lead to a valid polycube.}
\label{fig:invalidities}
\end{figure}

We thus define the \textbf{validity proxy} of a labeling $V_p(\ell)$ as:
\begin{equation} \label{eq:validity_def}
V_p(\ell) = |\mathcal{V}_{\text{inv}}| + |\mathcal{B}_{\text{inv}}| + \sum_{c \in \mathcal{C}_{\text{inv}}} (4 - N_c)
\end{equation}
where $N_c$ is the number of neighbors of chart $c$. Examples of labelings with non-empty invalid sets are shown in Figure \ref{fig:invalidities}. 

\noindent
A labeling $\ell$ will be considered \textit{pseudo-valid} if and only if $V_p(\ell)$ is equal to~0, or equivalently, the three invalid sets are empty. We provide some additional insight into the limitations of our validity proxy in Appendix \ref{app:validity}.

\textbf{Optimization criteria.}
Since our objective is the minimization of volumetric polycube parameterization distortion, which is too time-consuming to evaluate directly, our method requires a set of metrics on $\ell$ that approximate the distortion on $(\mathcal{V}', \mathcal{T}_\Omega)$.

We observe that a surface polycube of the boundary $\partial\mathcal{T}_\Omega$ can be computed significantly faster than its volumetric counterpart. We thus define the \textbf{workability} of a labeling $E_W(\ell)$ using the distortion of the mapping from $(\mathcal{V}, \partial\mathcal{T}_\Omega)$ to a fast polycube $(\mathcal{V}_{f}, \partial\mathcal{T}_\Omega)$. The term workability refers to the ability of some material to be easily deformed into a different shape. 
We compute our fast surface polycube as follows. Given a labeling $\ell$ defining a set of charts $\mathcal{C}$, we use a change of variables to constrain all vertices on a chart to the same value on the chart's label axis. We then minimize a least-squares problem aiming to preserve edge lengths on the two other axes. The solution yields a set of new vertex positions $\mathcal{V}_{f}$.
The per-triangle distortion is measured using the singular values $\sigma_1$ and $\sigma_2$ of the Jacobian of the mapping from the initial triangle to its equivalent in $(\mathcal{V}_{f}, \partial\mathcal{T}_\Omega)$ \cite{tarini2004polycube}:
$$
e_w = \sigma_1 + \sigma_2 + \frac{1}{\sigma_1 \sigma_2} + \frac{\sigma_1}{\sigma_2} + \frac{\sigma_2}{\sigma_1} - 4
$$

\begin{figure*}[tbp]
  \centering
  \mbox{} \hfill
  \includegraphics[width=1.0\linewidth]{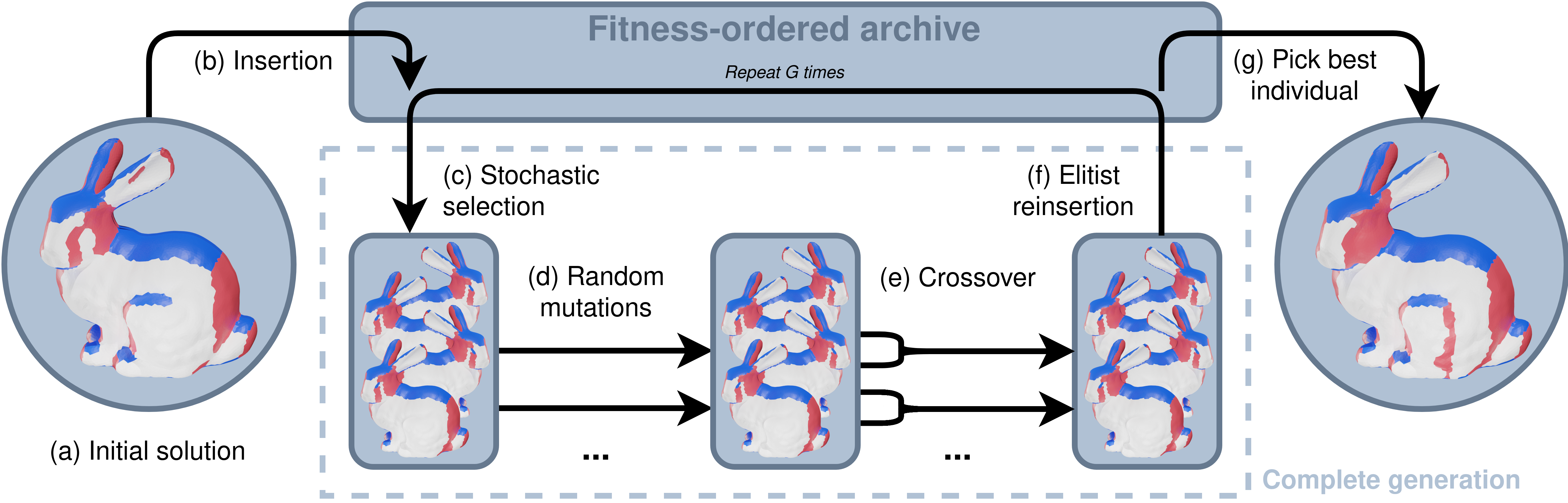}
  \caption{\label{fig:genetic}
           Architecture of our genetic optimization framework.}
\end{figure*}

For some invalid labelings, some triangles may be degenerate and have infinite $e_w$. We clamp $e_w$ to some arbitrarily large value, greatly penalizing invalid labelings. Finally, we compute the overall workability $E_W(\ell)$ by integrating $e_w^2$ over the boundary $\partial\mathcal{T}_\Omega$.  
The resulting surface polycube may contain inverted triangles which would require more time-consuming optimization to resolve, but in practice it remains a good proxy of the final polycube quality issued from $\ell$.

We complement our novel metric with additional ones used in previous work \cite{polycut_livesu}. First, the \textbf{fidelity} of a given face measures how well its label fits its orientation. It is defined as the dot product of its normal vector and its assigned direction. The fidelity of a labeling $E_F$ is then computed by integrating over all boundary triangles. Finally, \textbf{compactness} $E_C$ is the number of corner vertices in the polycube graph associated with $\ell$. 


By combining the above metrics, 
we define a $fitness$ function that can be embedded in our genetic framework:
$$
fitness(\ell) = V_p(\ell) + \omega_1 E_{W}(\ell) + \omega_2 E_F(\ell) + \omega_3 E_C(\ell)
$$

The coefficients vector $\omega = (\omega_1, \omega_2, \omega_3)$ is set to $(10^2, 10^{-2}, 10^{-2})$ in our experiments. This $fitness$ function is able to discriminate between candidate solutions and favor search directions reducing parameterization distortion as evaluated by our metrics.  



\subsection{Initial solution} \label{sec:init_sol}


Our algorithm requires an initial labeling to improve upon. We use the graph-cut initialization method presented by Livesu et al. \cite{polycut_livesu}. Considering the dual graph of a triangle mesh, its main advantage is the trade-off between:
\begin{itemize}
    \item a unary cost for assigning a given label to a face, penalizing label directions poorly aligned with triangle normals;
    \item and a binary cost for two neighboring triangles labeled differently, with a greater cost for neighbors that are close to be coplanar.
\end{itemize}


The unary cost helps find a labeling with low fidelity error, while the binary improves compactness and reduces boundary size. 
This method provides us with a satisfying initial solution that encompasses important fidelity information. While it is a good starting point, in general this solution has invalid patches. It cannot be used as is, and will be enhanced by our proposed genetic framework.
We use a graph-cut optimization library by Boykov et al. \cite{gco} which implements the improvements described in additional research \cite{gco2, gco3}. We allow neighbors with opposite orientations in our initial solution and propose our own fix in Section \ref{sec:repairs}.

\begin{figure}[htb]
\subfloat[Axis]{\includegraphics[width=0.11\hsize]{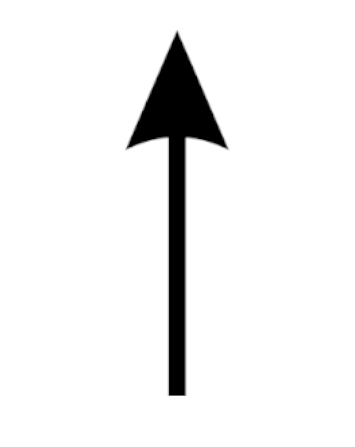}\label{fig_tp1}}\hfill
\subfloat[Boundary]{\includegraphics[width=0.2\hsize]{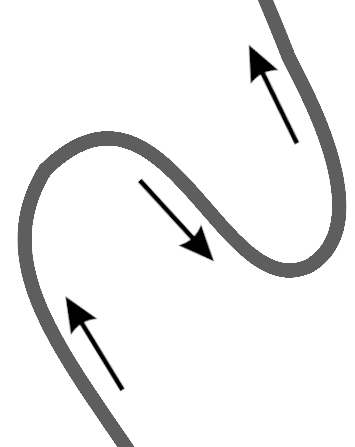}\label{fig_tp2}}\hfill
\subfloat[Unary bias]{\includegraphics[width=0.2\hsize]{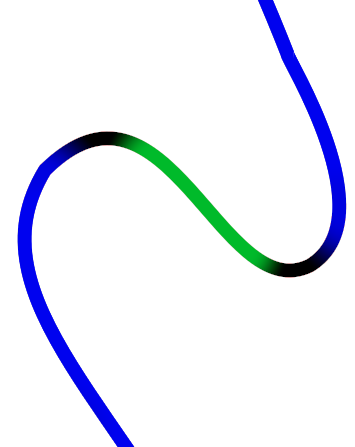}\label{fig_tp3}}\hfill
\subfloat[Binary cost]{\includegraphics[width=0.2\hsize]{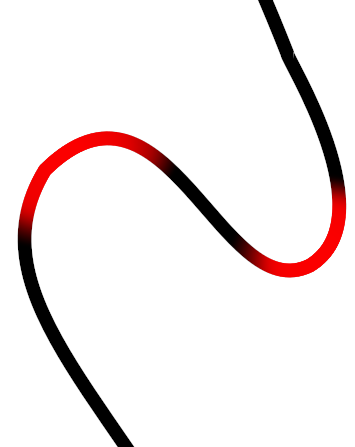}\label{fig_tp4}}\hfill
\subfloat[Labels]{\includegraphics[width=0.2\hsize]{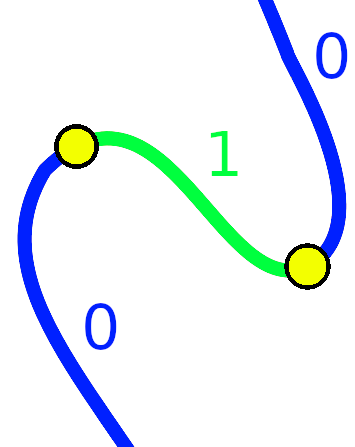}\label{fig_tp5}}
\caption{Turning points identification.}
\label{fig:turningpoints}
\end{figure}

We compute \textbf{turning points} following a graph-cut approach on boundary edges, similarly to Livesu et al. \cite{polycut_livesu}. Considering a directed boundary (Fig \ref{fig_tp2}) and its corresponding axis (Fig \ref{fig_tp1}), our graph-cut approach partitions the boundary with two labels (Fig \ref{fig_tp5}) and we define turning points as the vertices where the label switches. We use the unary cost $u$ (Fig \ref{fig_tp3}) to infer a bias based on the dot product of the axis and a given edge. Then, the binary cost $b$ encourages cuts between poorly aligned consecutive edges. Specifically, for a normalized edge $\vec{e}$ and its label $l_e$, we define these costs as follows:
\[
\begin{cases}
   u(\vec{e}, 0) = 1 - e^{-\frac{1}{2}(\frac{\vec{e} . \vec{axis}}{0.9})^2} \text{, if } \vec{e} . \vec{axis} < 0 \text{, and 0 otherwise};\\
   u(\vec{e}, 1) = 1 - e^{-\frac{1}{2}(\frac{\vec{e} . \vec{axis}}{0.9})^2} \text{, if } \vec{e} . \vec{axis} > 0 \text{, and 0 otherwise};\\
   b(\vec{e_1}, \vec{e_2}, l_{e_1}, l_{e_2}) =  e^{\frac{-(\vec{e_1} . \vec{e_2} - 1)^2}{2}} \text{ if }  l_{e_1} \neq l_{e_2} \text{, and 0 otherwise.}
\end{cases}
\]

\subsection{Genetic framework}


From an initial labeling, our framework generates new solutions and selects promising search directions. The overall architecture of our Evocube framework is illustrated in Figure \ref{fig:genetic}.
We define the key elements to select and cross individuals in this section, and labeling mutations in section \ref{sec:mutations}.

\textbf{Crossover.} The crossover operator takes two solutions $\ell_1$ and $\ell_2$ as input, and generates a new individual including both of its parents' mutations. To achieve this, when both parent solutions agree on a face label, the child is assigned the same label. When labels conflict, the label that was changed during the most recent generation is kept. In case of a draw, we consistently pick $\ell_1$'s label.

\textbf{Archive.} We use an archive system to keep track of the best solutions to date. The archive has a fixed maximum size, and when an element is submitted, sufficient $fitness$ greater than that of the worst element is required. The latter is then discarded.
When selecting an element from the archive, we use a stochastic model favoring higher ranked solutions, similarly to Lim et al. \cite{LIM2020}. For an archive containing ranked solutions from $\ell_1$ to $\ell_n$, the probability of randomly picking $\ell_i$ is as follows:
$$
P(\ell_i) = \frac{n - i + 1}{1 + ... + n}
$$

\textbf{Generations.}
Here, we describe the core architecture of our genetic framework illustrated in Figure \ref{fig:genetic}. The initial solution is generated using the graph-cut approach described previously. 
At the beginning of each generation, $N$ candidate solutions are selected from the archive. Following our stochastic model, the highest-ranked solutions are picked several times. The selected individuals then undergo random mutations. Since individual mutations are independent, they are computed in parallel and a greater $N$ can be used while maintaining reasonable time complexity. \\ \noindent
Afterwards, pairs of individuals are stochastically selected and combined using the crossover operator. This is repeated $C$ times, leading to a population of size $N+C$. Finally, we insert all new solutions with sufficient score in the archive, and discard all unfit individuals, ending the generation. 

This process ends after $G=40$ generations, or after three consecutive generations without change in the archive's best solution. In our experiments, we found that $N=100$ and $C=10$ were sufficient to reach convergence for most input meshes.

\subsection{Labeling mutations}\label{sec:mutations}

\begin{figure}[tb]
\subfloat[Directional path]{\includegraphics[width=0.33\hsize]{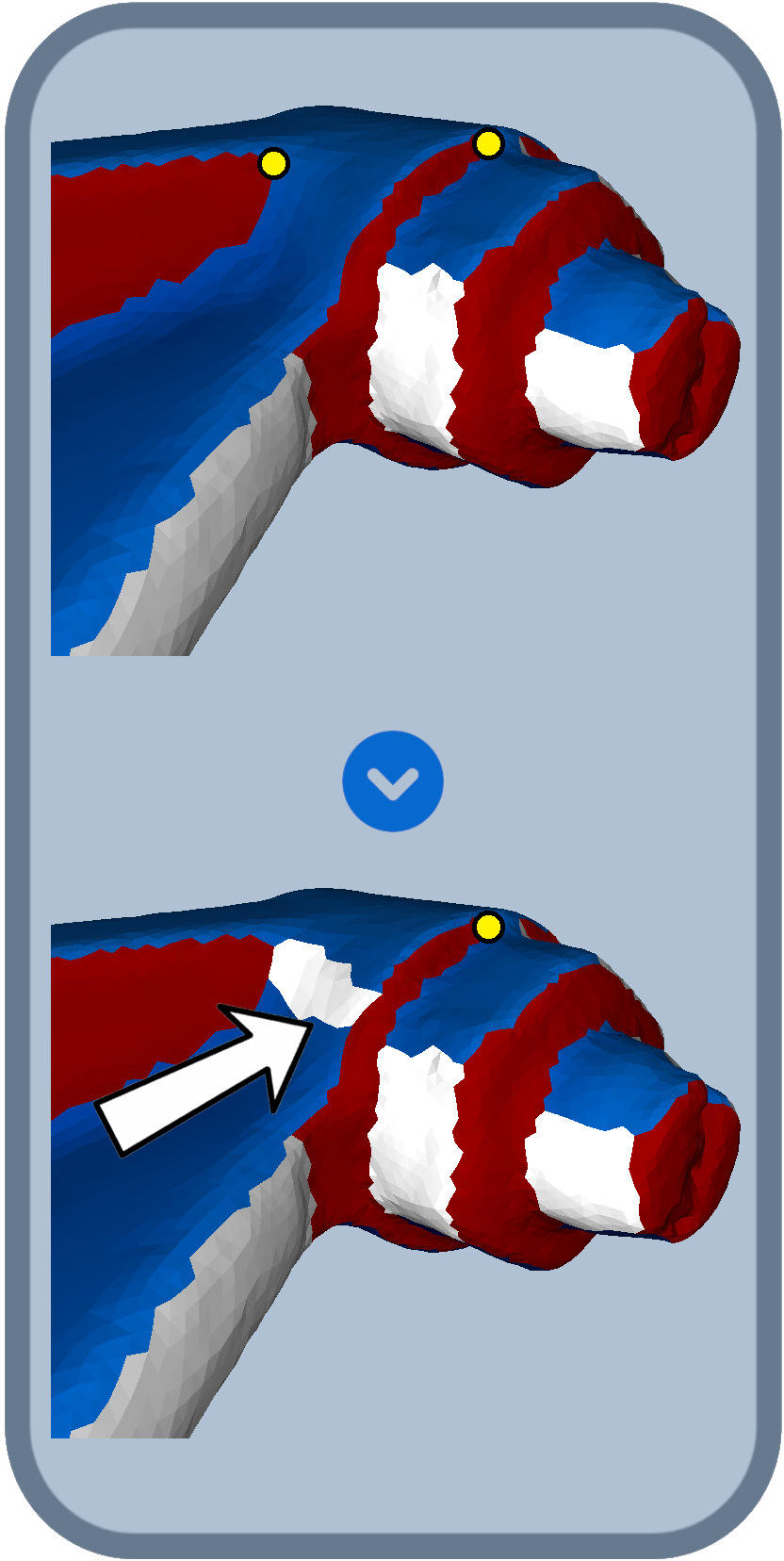}\label{fig_mut1}}\hfill
\subfloat[Chart removal]{\includegraphics[width=0.33\hsize]{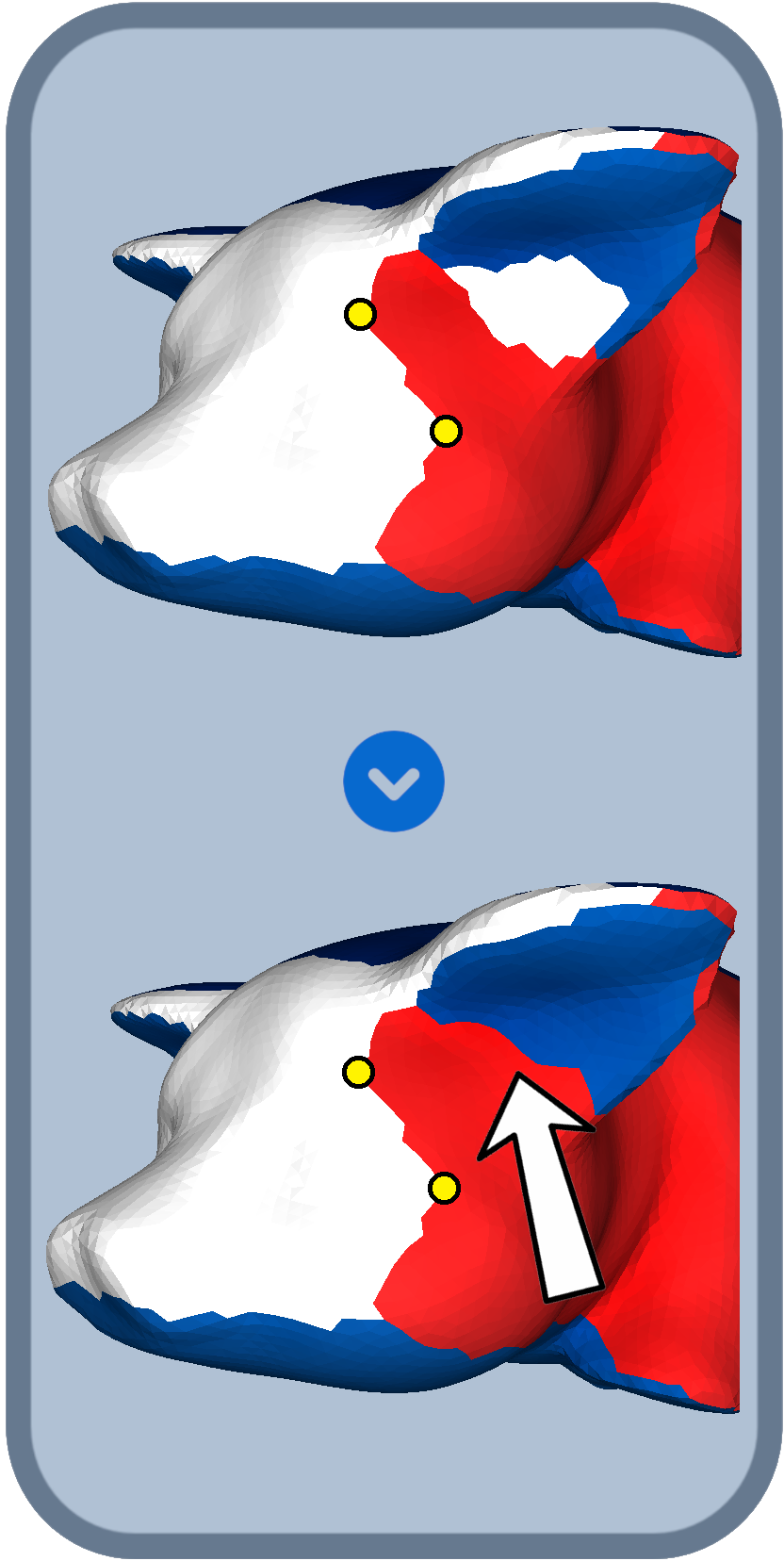}\label{fig_mut2}}\hfill
\subfloat[Chart propagation]{\includegraphics[width=0.33\hsize]{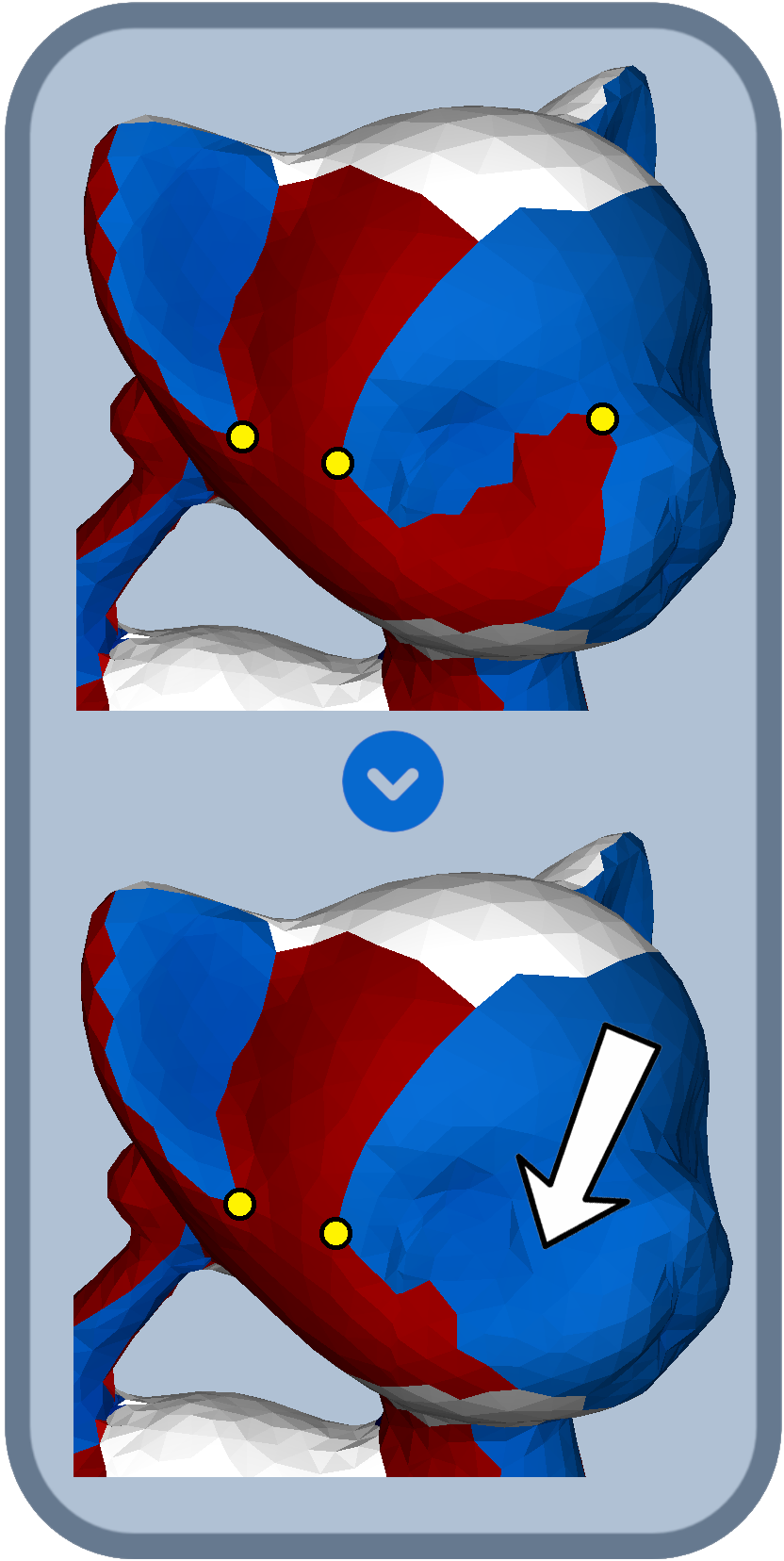}\label{fig_mut3}}
\caption{Base mutations in our genetic framework. Turning points are highlighted in yellow.}
\label{fig:mutations}
\end{figure}

We propose a set of mutations that aim to improve an initial labeling. Given that the search space of all possible labelings is of size~$6^{|\partial \mathcal{T}_\Omega|}$, we restrict our search to meaningful solutions. Our mutations specifically focus on fixing invalid patches and identified turning points. These operations may yield poor results under some circumstances, but by virtue of our genetic framework, our method identifies and discards detrimental modifications. Our labeling mutations are illustrated in Figure \ref{fig:mutations}. Future work may complement \textit{Evocube} with additional operations to account for specific configurations, but we found these to be sufficient for most geometries.


\textbf{Directional path.} In Figure \ref{fig_mut1}, a turning point is chosen randomly, along with one of the 4 orthogonal directions on the chart. We then greedily select a path matching that desired direction until we reach another boundary. Once the path is defined, we apply the new label to the triangles around the path and propagate to their neighbors. The introduced label is picked from only two possibilities, since the chart's label and its opposite cannot be picked, and the path is highly non-constant on its propagation axis, effectively forbidding two more labels. This operation is similar to the orthogonal path described by Gregson et al. \cite{gregson_polycube_2011}.



\textbf{Chart removal.} When a chart is invalid as in Figure \ref{fig_mut2}, it can sometimes simply be removed. To perform this, we use the same graph-cut method used to compute an initial solution, with two modifications. Firstly, the rest of the labeling is locked to prevent any unwanted modification, and secondly we forbid the initial label to be assigned again to the chart that is being removed.


\textbf{Chart Propagation.} Given a random border with a turning point, the label of one side is applied to the other side around the selected turning point, as illustrated in Fig \ref{fig_mut3}. If the border is already monotone, then we propagate along the whole border instead.

In our genetic framework, mutations are chosen randomly and applied in random locations. In order to help speedup convergence, we select among invalid charts if some remain for chart removal, or among turning points for directional path and chart propagation. If none remains, the location is randomly selected among all charts, or boundary vertices, respectively. In our experiments, the distance for propagation mutations is sampled from $[l_{avg},\text{ 5 }l_{avg}]$, where $l_{avg}$ is the average edge length in the mesh.

\subsection{Labeling repairs}\label{sec:repairs}

\begin{figure}[tb]
\subfloat[Opposite boundaries]{\includegraphics[width=0.33\hsize]{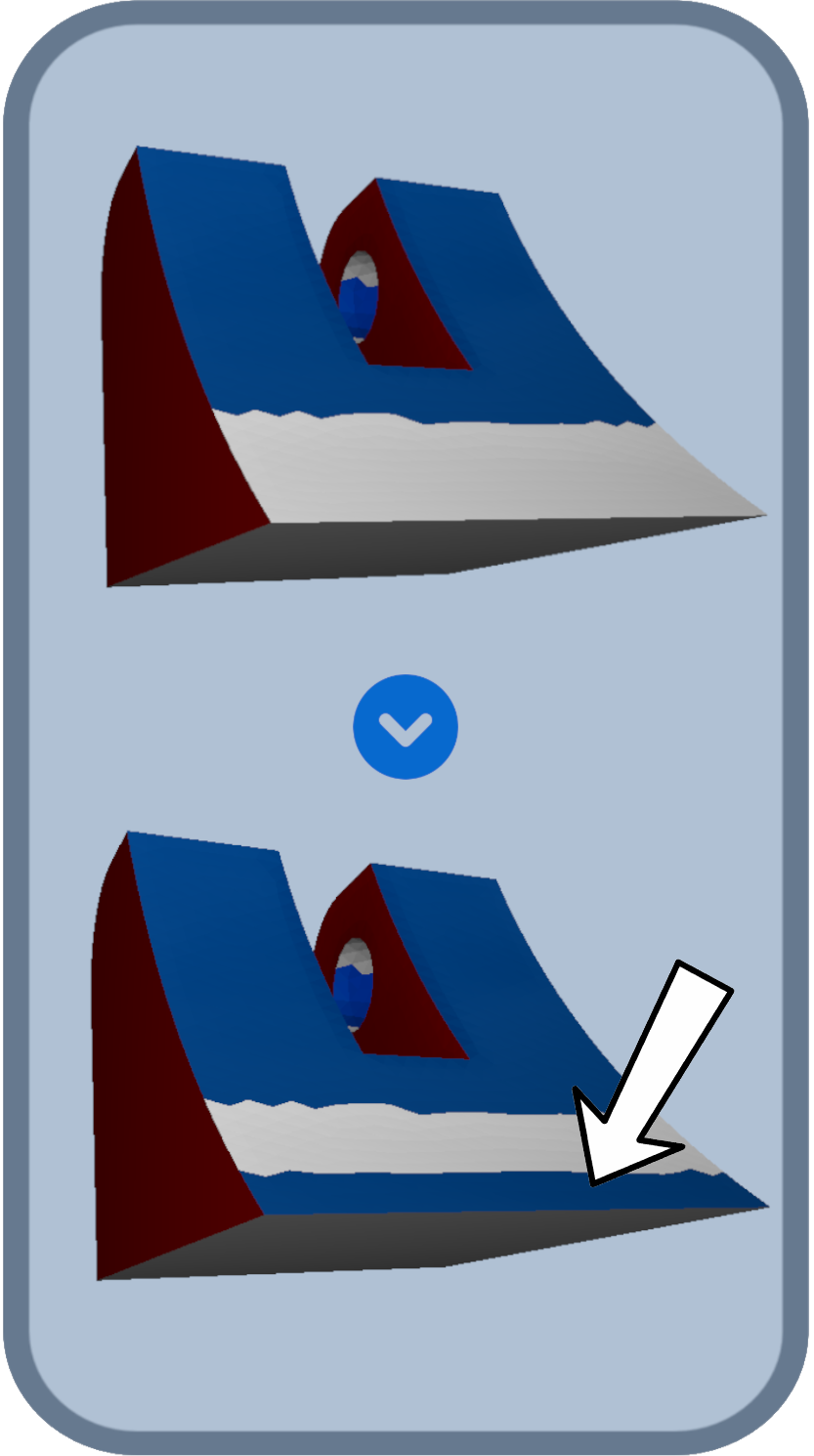}\label{fig_rep1}}\hfill
\subfloat[High valency corner]{\includegraphics[width=0.33\hsize]{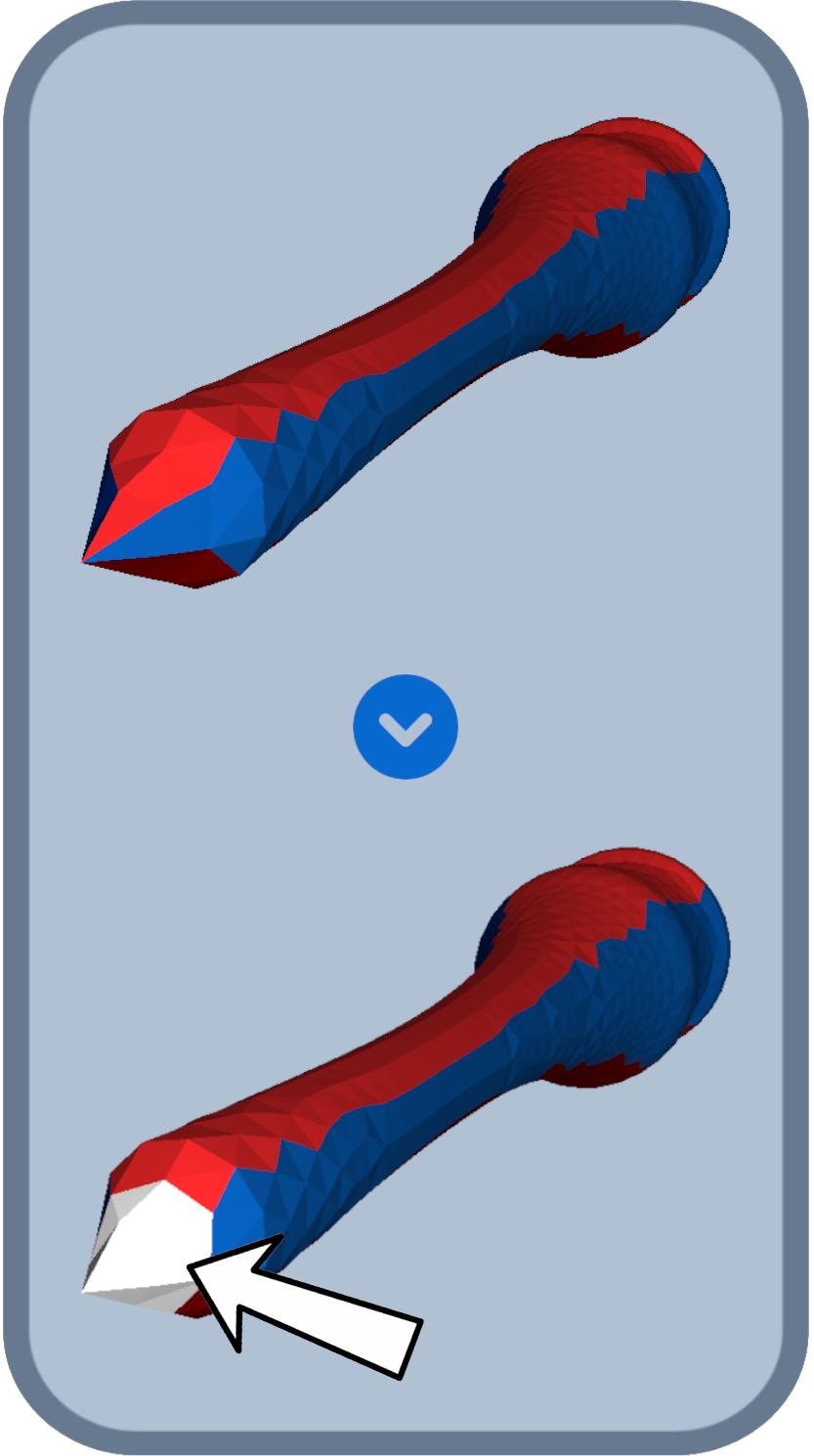}\label{fig_rep2}}\hfill
\subfloat[Path smoothing]{\includegraphics[width=0.33\hsize]{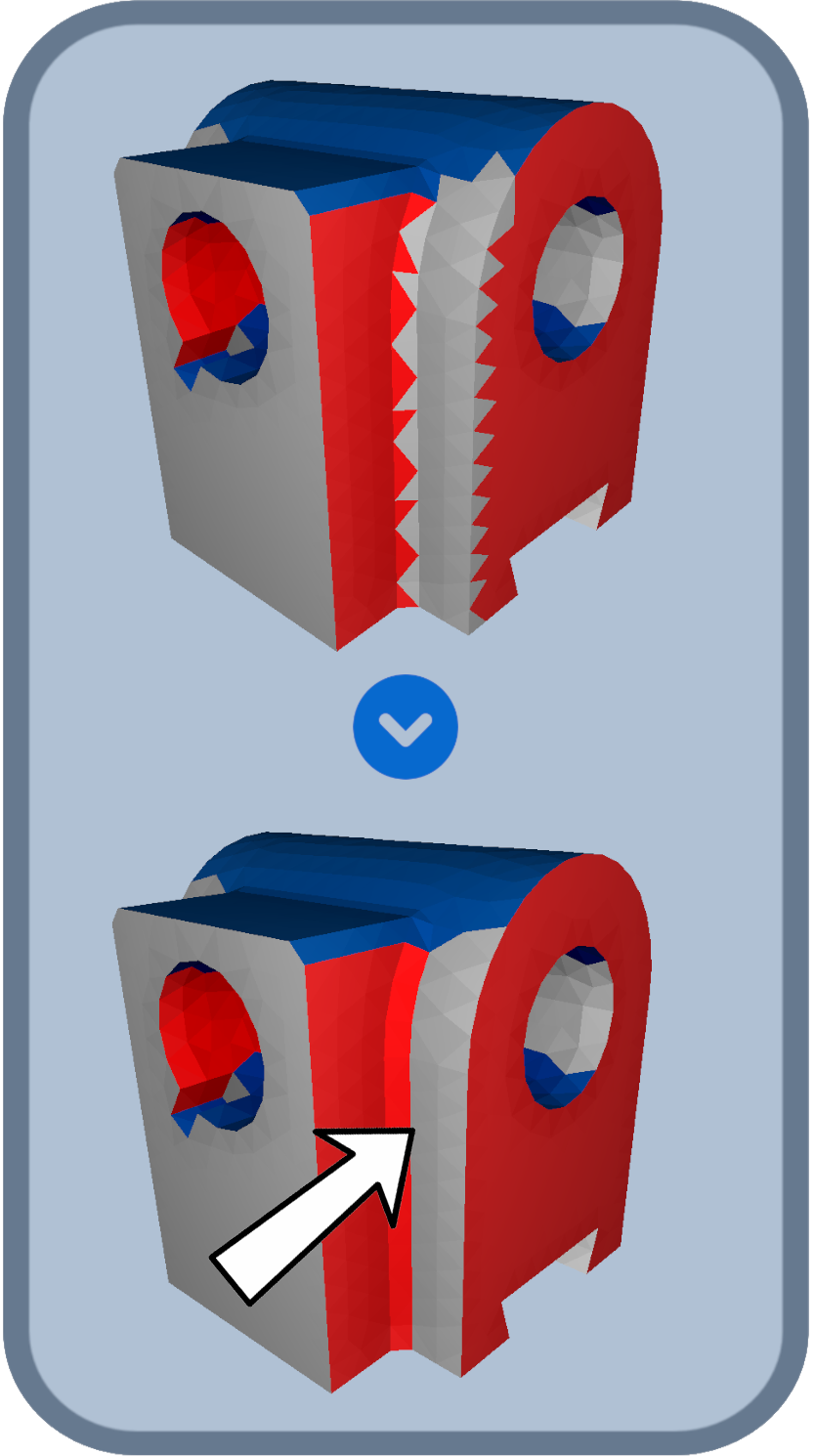}\label{fig_rep3}}
\caption{Labeling repairs.}
\label{fig:labeling_repairs}
\end{figure}

We complement our labeling mutations with a set of fixes illustrated in Figure \ref{fig:labeling_repairs} which rectify some easily identifiable labeling problems. Unlike mutations, these operations are applied deterministically, as specified below. 


In a valid labeling, no pair of neighboring charts can have opposite labels. We repair any boundary separating two such charts by introducing a new chart with one of the remaining labels around the boundary (Fig \ref{fig_rep1}). The new chart can be inserted on both sides of the boundary or on either side alone, and its size is taken to be a multiple of the mesh's average edge length. We pick the preferred option by measuring $fitness$ for all possibilities. We apply this repair only on the initial and final solutions.


Similarly, we propose a simple fix for corners with invalid valency in the polycube graph (Fig \ref{fig_rep2}). We introduce a new chart with one of the remaining labels around the problematic vertex, and pick the optimal size using our $fitness$ evaluator. Similarly, this operation is used only on the initial and final solutions.

Finally, any triangle with two edges on the same chart boundary will be flattened in parameterization space. We consider pairs of triangles to be neighbors only if they share an edge on a chart boundary. We then smooth boundaries by relabeling any triangle surrounded by two neighbors sharing a label different of its own (Figure \ref{fig_rep3}). In some cases, several iterations are needed to obtain a smooth labeling. This operation is fast and we apply it to any individual before evaluation.



%% file: p5-polycube-maps.tex
\section{Hexahedral meshing} \label{sec:def_extr}

Given a pseudo-valid polycube labeling $\ell$ of the boundary mesh $(\mathcal{V}', \partial\mathcal{T}_\Omega)$, we now seek to compute a volumetric polycube-map matching this labeling and an all-hex mesh. Note that previous work \cite{gregson_polycube_2011,polycut_livesu} has also described effective methods which could be used in combination with Evocube.

We compute a low-distortion volumetric polycube of $(\mathcal{V}, \mathcal{T}_\Omega)$. This differs from the polycube-based fitness evaluator described in Section \ref{sec:fitness}, which is only a fast surface mapping of $(\mathcal{V}, \partial\mathcal{T}_\Omega)$.
Similarly, we first proceed to a change of variable along chart vertices to enforce the polycube topology induced by our labeling $\ell$. Consequently, vertices on the same chart will share the same coordinate on the axis associated with the chart's label. We then minimize a Laplacian energy penalizing edge distortion over both interior and exterior vertices. 
After this step, the resulting polycube may still contain inverted tetrahedra. Subsequently, we minimize the energy proposed by Garanzha et al. \cite{garanzha2021foldover} under the same change of variable to restore inverted cells and further improve the quality of the volumetric polycube-map. 

For all-hex meshing, we then quantize our polycube-map using the robust mixed-integer method described by Protais et al. \cite{protais_robust_2020}, and extract our initial hexahedral mesh by applying the inverse mapping from our quantized polycube to input space. 
While polycube-based hexahedral meshing is appreciated for its regularity, it also suffers from the lack of inside singularities. 
We introduce a layer of elements on the boundary, and smooth the resulting hex mesh by minimizing a combination of the energy defined in \cite{branets_distortion_2002} and the mesh distance to the input boundary and features. 

There exists various methods to extract an hexahedral mesh from a valid labeling. We include our method for completeness. In practice, it has proven to be robust and effective during our experiments. To further improve on the quality of hex meshes extracted from polycubes and push singularities towards the interior, our simple pillowing can be replaced with more advanced methods such as selective-padding \cite{cherchi2019selective} or cut-enhancements \cite{guo2020cut}.

%% file: p6-experiments.tex
\section{Experiments} \label{sec:experiments} 

\begin{figure*}[tb]
  \centering
  \includegraphics[width=1.0\linewidth]{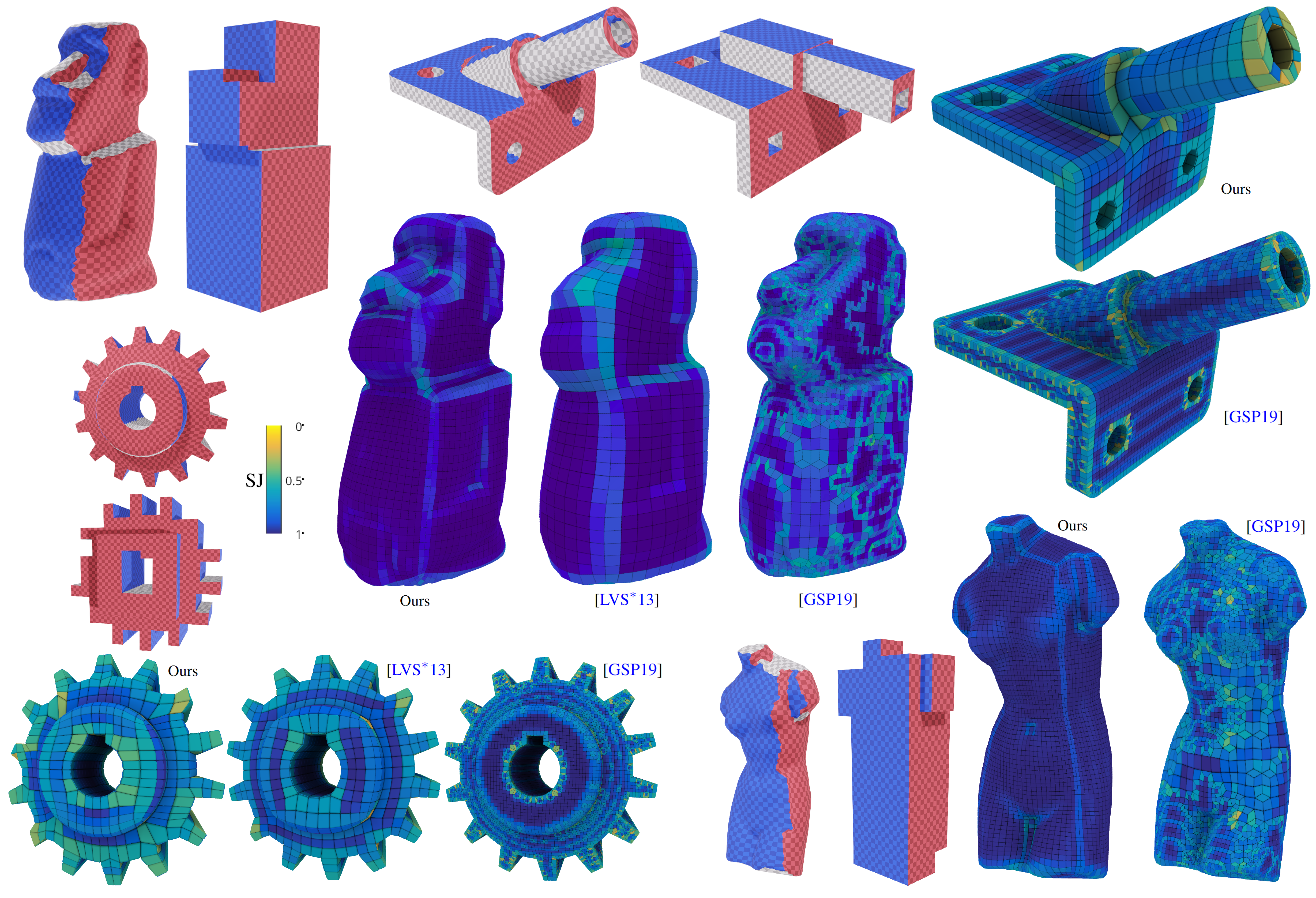}
  \caption{\label{fig:comparison_general}
           Comparison with previous methods on \textit{moai},  \textit{cami1},  \textit{pinion} and  \textit{venus}. Mesh quality is reported in Table \ref{tab:comp_results}.}
\end{figure*}


We tested our method on a collection of 1315 meshes from the MAMBO dataset \cite{MAMBODataset}, the natural and CAD shapes used by Gao et al. \cite{gao2019feature}, and a subset of ABC \cite{Koch_2019_CVPR_ABC} consisting of the inputs sampled by Reberol et al. \cite{ReberolQuasiStructured}. 
Our results on these datasets are synthesized in Table \ref{tab:dataset_results}.
We generate initial solutions with a ratio of ${\omega_\text{unary}}/{\omega_\text{binary}} = 3$, as recommended by Livesu et al. \cite{polycut_livesu}. If no pseudo-valid labeling is found on our first attempt, we divide this ratio by 3 and apply our labeling optimization starting from this new initial solution.

We generate tetrahedral meshes with \textit{netgen} \cite{netgen}, and use \textit{libigl} \cite{libigl} for common geometry processing operations. 
In the following, all-hex meshes are visualized in \textit{HexaLab} \cite{BRACCI201924_hexalab}. In our supplemental material, we provide renderings of all labelings and polycubes, as well as per-model statistics.

\begin{table}[htb]
\small
\centering
\begin{tabular}{|l|l|l|l|l|l|}
\hline
Dataset & size & mesh & $V_p(\ell) = 0$ & $D_a<2$ & minSJ$\geq 0$ \\ \hline \hline
Mambo &   113 &  100\%  & 100\%  & 97.4\%    &  96.5\% \\ 
\hline \hline
Smooth &  93 &  100\%   & 100\%     & 75.2\%    &  89.2\% \\
CAD &  109  &  100\% & 94.5\%     & 87.1\%    &  85.3\% \\ \hline 
\hline 
ABC &  1000 &  99.4\%   & 95.5\% & 82.3\%*    &  61.9\%* \\
\hline 
\end{tabular}
\caption{\label{tab:dataset_results}
           Results on CAD and smooth datasets. For each entry we report the number of inputs, the ratios of (a) tetrahedral meshes successfully generated, (b) pseudo-valid labelings, (c) polycubes with area distortion $D_a$ \cite{tarini2004polycube} below 2 (the ideal value being 1), and (d) hex meshes with minSJ$\geq 0$. *Given the size of the ABC dataset, these results were simply computed using our fast polycube estimator and \textit{libHexEx} \cite{lyon2016hexex} without any post-processing.}
\end{table}

\subsection{Comparison}

We compare our work with PolyCut \cite{polycut_livesu} using the binaries provided by the authors incorporating subsequent work \cite{Livesu:2015:Untangler}. On some inputs, we had to stop labeling optimization after one hour without progress. We also report on the initial solution computed with graph-cut, showing the importance of our labeling optimization. 
The results, detailed in Table \ref{tab:comp_results}, show that our method converges to a pseudo-valid labeling in an overwhelming majority of cases. 
When compared with another labeling based method \cite{polycut_livesu}, our method is more robust and our labelings subsequently lead to high quality hexahedral meshes in a greater range of cases. Our ability to compute satisfying polycubes on more models than previous work can be used to improve the robustness of previous polycube enhancement methods \cite{cherchi2019selective,guo2020cut} and further enhance our resulting hex meshes.

As explained in Section \ref{sec:def_extr}, we compute volumetric polycubes and as such we do not optimize for surface polycube distortion \cite{tarini2004polycube}. Our labelings and volumetric polycubes aim to pre-determine polycube topology and the number of cells subdividing the input volume. The exact position of these cells in the final hex mesh is determined by our all-hex smoothing, effectively removing the need for surface polycubes in comparison with previous work. 
To compute a low-distortion surface parameterization, our polycubes can be used as an initial solution for an inverse optimization with remeshing such as the one used by Livesu et al. \cite{polycut_livesu}. 

We also compare our results with the feature-aware octree-based method proposed by Gao et al. \cite{gao2019feature}, which is known to be more robust than polycubes but yield less regular meshes. As expected, our method produces satisfying hex meshes on a smaller range of models but still results in better average cell quality on the whole dataset. As illustrated in Figure \ref{fig:comparison_general}, our method produces highly regular hex-meshes with cells of similar sizes, a property which is greatly appreciated for the interpretation of numerical simulation results. In comparison, the results of Gao et al. \cite{gao2019feature} include cells of varying sizes and are able to capture complex shapes with more accuracy, at the expense of interpretability and singularities. 

\begin{table}[tb]
\small
\centering
\begin{tabular}{|l|l|l|l|l|}
\hline
& & & & \\[-1em] 
Method & pseudo-valid & minSJ$\geq 0$ & $\overline{\text{minSJ}}$  & $\overline{\text{avgSJ}}$\\ \hline \hline
Mambo Basic &   &     &    &\\ 
Graph-cut &  77.0\% & & &    \\
\cite{polycut_livesu} &  94.6\%  &  73.0\% & -0.00     & 0.60    \\ 
Ours &  \textbf{100\%}  &  \textbf{100\%} & \textbf{0.21}     & \textbf{0.90}    \\ \hline 
\hline 
Mambo Simple &   &     &    &\\ 
Graph-cut &  66.7\% & & &    \\
\cite{polycut_livesu} &  96.7\%  &  53.3\% & -0.24  & 0.40    \\ 
Ours &  \textbf{100\%}  &  \textbf{90.0\%} & \textbf{0.08}     & \textbf{0.84}   \\ \hline 
\hline 
Mambo Medium &   &     & &\\ 
Graph-cut &  33.3\% & & &    \\
\cite{polycut_livesu} &  88.9\%  &  11.1\% & -0.54 & 0.06  \\ 
Ours &  \textbf{100\%}  &  \textbf{88.9\%} & \textbf{-0.00} & \textbf{0.75}  \\ \hline 
\hline 
Smooth \cite{gao2019feature}  &  &  & &\\
Graph-cut &  18.3\% & & &    \\
\cite{polycut_livesu} &  76.1\%  &  55.7\% & -0.20     & 0.37    \\ 
\cite{gao2019feature} &  N/A  &  \textbf{100\%} & \textbf{0.19}     & 0.81    \\
Ours &  100\%  &  89.2\% & 0.15     & \textbf{0.86}    \\ 
\hline 
\hline 
CAD \cite{gao2019feature}  &  &  & &\\
Graph-cut &  72.5\% & & &    \\
\cite{polycut_livesu} &  85.8\%  &  52.8\% & -0.23     & 0.39   \\ 
\cite{gao2019feature} &  N/A  &  \textbf{100\%} & 0.17     & 0.83    \\ 
Ours &  94.5\%  &  85.3\% & \textbf{0.19}     & \textbf{0.87}    \\ 
\hline 
\end{tabular}
\caption{\label{tab:comp_results}
           Comparison in terms of ratio of pseudo-valid labelings, hex meshes with minSJ$\geq 0$, and hex quality. We denote $\overline{X}$ the average of $X$ over a given dataset. When no output is produced, the worse possible value of $-1$ is assumed instead.}
\end{table}

\begin{table}[tb]
\small
\centering
\begin{tabular}{|l|l|l|l|l|l|}
\hline
Input & $N_c$ & angle/area d. & minSJ  & avgSJ & \#hex \\
\hline \hline
Rocker Fig \ref{fig:pipelineOverview} & & & & &    \\
\cite{gao2019feature} &    & & 0.165 & 0.813 & 30k \\ 
\cite{polycut_livesu} &  62  & 1.066/1.051 & \textbf{0.370} & 0.890 & 57k \\ 
Ours &  74  & 1.315/1.523 & 0.241    & \textbf{0.938} & 60k \\
\hline \hline
Moai Fig \ref{fig:comparison_general} & & & & &    \\
\cite{gao2019feature} &    & & 0.263 & 0.824 & 25k \\ 
\cite{polycut_livesu} &  8  & & 0.497 & 0.950 & 3k \\ 
Ours &  28  & 1.073/1.057 & \textbf{0.530}     & \textbf{0.968} & 18k \\ 
\hline 
cami1 Fig \ref{fig:comparison_general} & & & & &    \\
\cite{gao2019feature} &    & & 0.056 & \textbf{0.829} &  25k \\ 
Ours &  72  & 1.164/1.221 & \textbf{0.108}     & 0.802 & 3k \\
\hline 
pinion Fig \ref{fig:comparison_general} & & & & &    \\
\cite{gao2019feature} &    & & 0.026 & 0.783 & 89k   \\ 
\cite{polycut_livesu} &    & & 0.106 & \textbf{0.863}  & 5k  \\ 
Ours & 112 & 1.122/1.096 & \textbf{0.142}     & 0.816 & 5k  \\
\hline 
venus Fig \ref{fig:comparison_general} & & & & &    \\
\cite{gao2019feature} &    & & 0.131 & 0.800 &  20k \\ 
Ours &  36  & 1.131/1.122 & \textbf{0.326}     & \textbf{0.950} & 49k \\
\hline 
\hline 
S22 Fig \ref{fig:bad_feature_edge} & 64 & 1.053/1.032 & 0.023 & 0.910 & 23k \\
\hline 
\hline
S26 Fig \ref{fig:mambo_results} & 80 & 1.044/1.027 & 0.016 & 0.899 & 21k\\
B16 Fig  & 16 & 1.160/1.161 & 0.098 & 0.917 & 10k\\
B51 Fig  & 32 & 1.115/1.101 & 0.085 & 0.951 & 20k\\
B76 Fig  & 26 & 1.098/1.058 & 0.059 & 0.934 & 8k\\
B38 Fig  & 24 & 1.303/1.149 & 0.058 & 0.845 & 9k\\
B49 Fig  & 12 & 1.020/1.021 & 0.083 & 0.989 & 49k\\
\hline
\end{tabular}
\caption{\label{tab:hex_meshes}
           Statistics of presented polycubes and hex meshes. We report the number of corners $N_c$ of polycubes, angle and area distortion, as well as quality metrics and number of cells for hex meshes.}
\end{table}

\subsection{Analysis}
 
Owing to its genetic nature, Evocube is highly parallelizable and time-efficient. Each generation, individuals mutate and are evaluated independently. In our experiments, this led to a speedup by a factor of 14, as illustrated in Figure \ref{fig:time_plot}. Despite our fast surface polycube evaluator, which is several magnitudes faster than computing the final polycube, the main time bottleneck of our framework remains fitness evaluation. On average, our method required around 15 seconds per input for labeling optimization and was able to compute the labelings leading to Table \ref{tab:dataset_results} in under six hours, as detailed in Appendix \ref{app:time}.

\definecolor{preliminary}{HTML}{dfff00}%
\definecolor{createindiv}{HTML}{ffbf00}%
\definecolor{mutations}{HTML}{ff7f50}%
\definecolor{chartscolor}{HTML}{de3163}%
\definecolor{fitnesscolor}{HTML}{9fe2bf}%
\definecolor{crossingcolor}{HTML}{40e0d0}%
\definecolor{archivecolor}{HTML}{6495ed}%
\definecolor{postcolor}{HTML}{ccccff}%
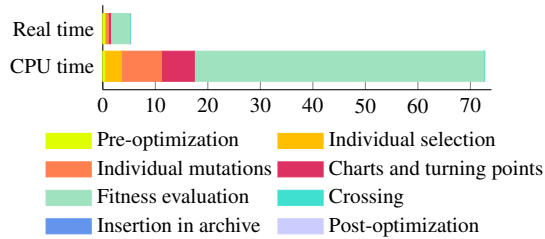
\begin{figure}%
\centering%
\begin{tikzpicture}%
\begin{axis}[%
    xbar stacked,%
    legend style={%
        legend columns=2,%
        at={(xticklabel cs:0.5)},%
        anchor=north,%
        draw=none,%
        cells={anchor=west}, 
        align=left%
    },%
    ytick=data,%
    axis y line*=none, 
    axis x line*=bottom,%
    tick label style={font=\footnotesize},%
    legend style={font=\footnotesize},%
    label style={font=\footnotesize},%
    xtick={0,10,20,30,40,50,60,70},%
    width=0.8\linewidth,%
    bar width=4mm,%
    xlabel={Time in hours},%
    yticklabels={CPU time, Real time},%
    xmin=0,%
    xmax=74,%
    area legend,%
    y=5mm,%
    enlarge y limits={abs=0.625},%
]%
\addplot[preliminary,fill=preliminary] coordinates%
{(0.330589,0) (0.330589,1)};%
\addplot[createindiv,fill=createindiv] coordinates%
{(3.20237,0) (0.212645,1)};%
\addplot[mutations,fill=mutations] coordinates%
{(7.63491,0) (0.506977,1)};%
\addplot[chartscolor,fill=chartscolor] coordinates%
{(6.25214,0) (0.415157,1)};%
\addplot[fitnesscolor,fill=fitnesscolor] coordinates%
{(55.1577,0) (3.66261,1)};%
\addplot[crossingcolor,fill=crossingcolor] coordinates%
{(0.00266842,0) (0.00017719,1)};%
\addplot[archivecolor,fill=archivecolor] coordinates%
{(0.000410876,0) (2.72831e-05,1)};%
\addplot[postcolor,fill=postcolor] coordinates%
{(0.205631,0) (0.205631,1)};%
\legend{Pre-optimization, Individual selection,  Individual mutations, Charts and turning points, Fitness evaluation, Crossing, Insertion in archive, Post-optimization}%
\end{axis}  %
\end{tikzpicture}%
\caption{Time plot (in hours of real and CPU time) of our labeling optimization over all 1315 tested models. We provide a detailed time report in Appendix \ref{app:time}.}%
\label{fig:time_plot}
\end{figure}%

\begin{figure}[tb]
  \centering
  \includegraphics[width=1.0\linewidth]{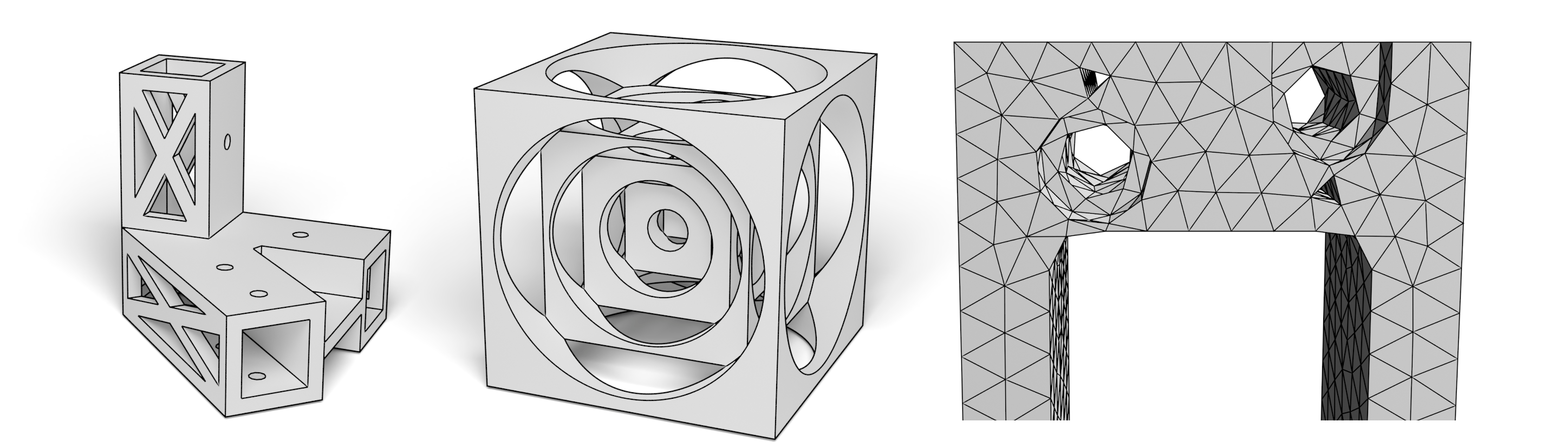}
  \caption{\label{fig:invalid_fail}
            Example meshes for which our Evocube implementation did not converge to a pseudo-valid labeling. The majority of failures fell in one of these categories: complex geometries with features poorly aligned with the base axes (left); shapes requiring a polycube with valency 4 corners (middle); and inputs with tiny holes that are too coarse to be satisfyingly labeled (right).}
\end{figure}

Nonetheless, our method failed to find valid labelings on some inputs. Similarly to previous research on polycube-maps, our method is unable to deal with complex shapes when they are poorly aligned with principal axes, as illustrated in Figure \ref{fig:invalid_fail}.
There are also cases in which a pseudo-valid labeling did not lead to a satisfying hex mesh, as reported in Tables \ref{tab:dataset_results} and \ref{tab:comp_results}, where the ratio of such labelings is higher than the ratio of successful hex-meshes.

\begin{figure*}[tb]
  \centering
  \includegraphics[width=1.0\linewidth]{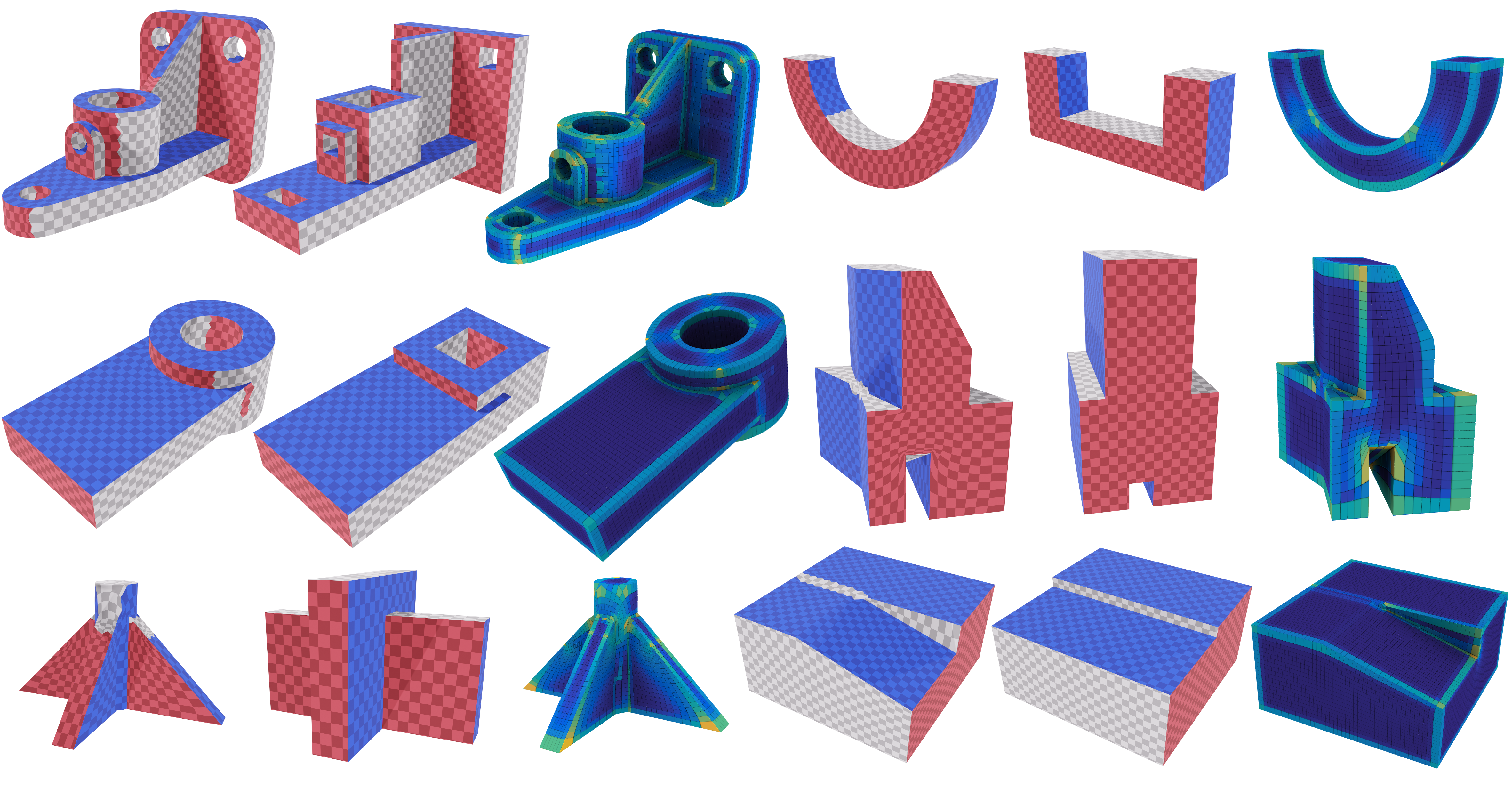}
  \caption{\label{fig:mambo_results}
           Additional results on CAD inputs from the MAMBO dataset. From left to right, top to bottom: S26, B16, B51, B76, B38, B49.}
\end{figure*}

By construction, our labeling corresponds to base axes and is dependent on the input's orientation. For some shapes, this assumption leads to non-optimal labelings. For example, a U-shaped cylinder will be mapped to a U-shaped polycube, as illustrated in Figure \ref{fig:mambo_results}, whereas a simple cuboid would be preferable and lead to fewer singularities in the resulting hex-mesh. This limitation could be addressed in our framework with recent work focusing on removing undesired polycube corners \cite{MANDAD2022102078}.

\begin{figure*}[tb]
  \centering
  \includegraphics[width=1.0\linewidth]{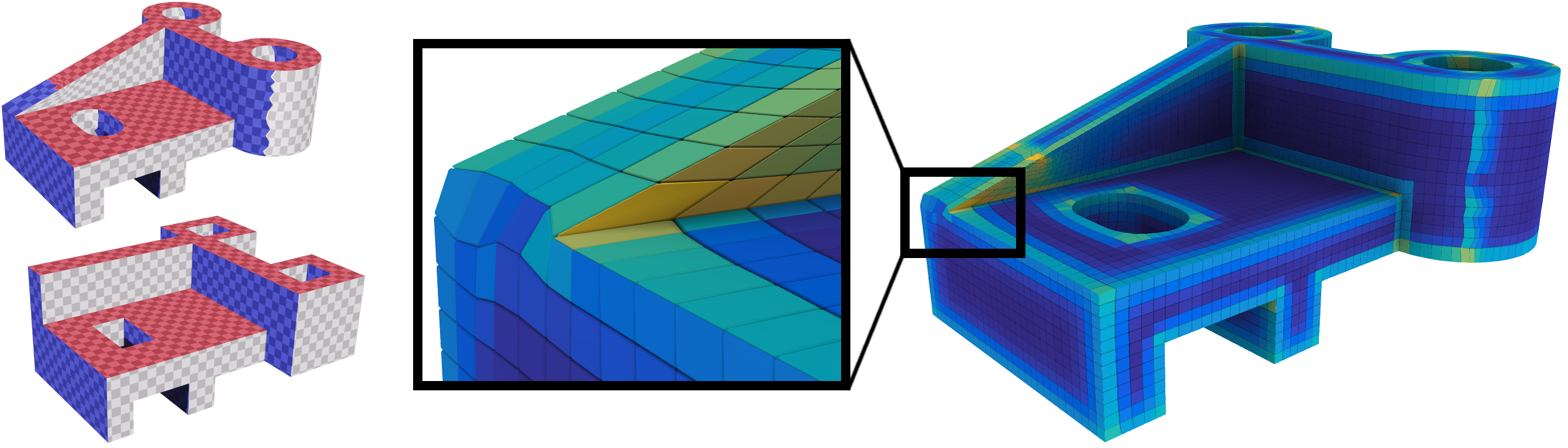}
  \caption{\label{fig:bad_feature_edge}
           In some cases, Evocube finds a low-distortion labeling that assigns the same label across a feature edge. In our current post-processing pipeline and similarly to other polycube-based methods, such feature edges are not preserved.}
\end{figure*}

\begin{figure}[htb]
\subfloat[Initial]{\includegraphics[width=0.3\hsize]{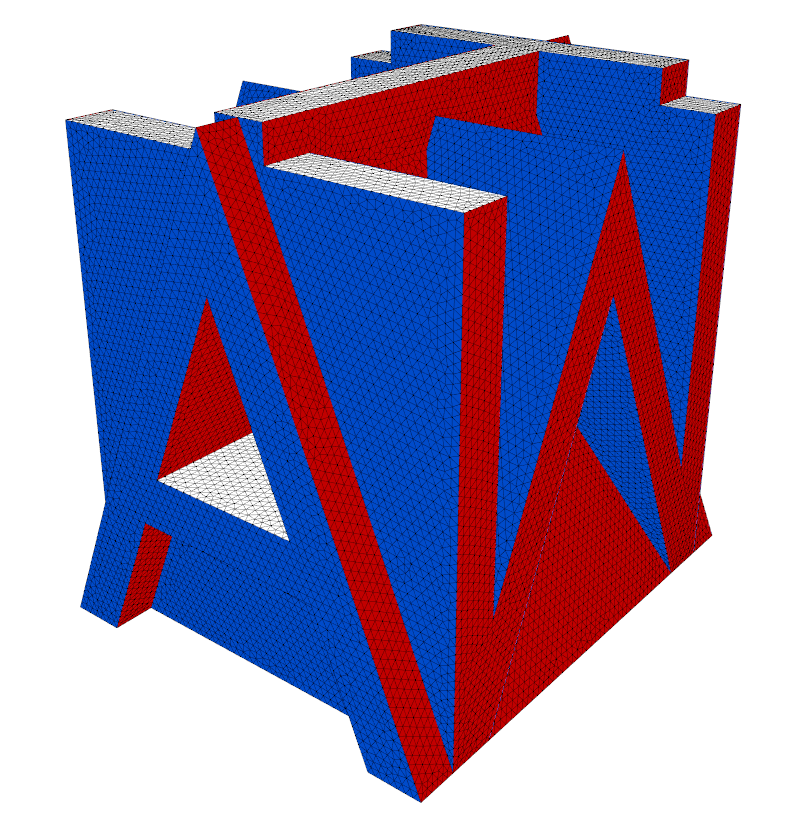}\label{fig_lt1}}\hfill
\subfloat[5 generations]{\includegraphics[width=0.3\hsize]{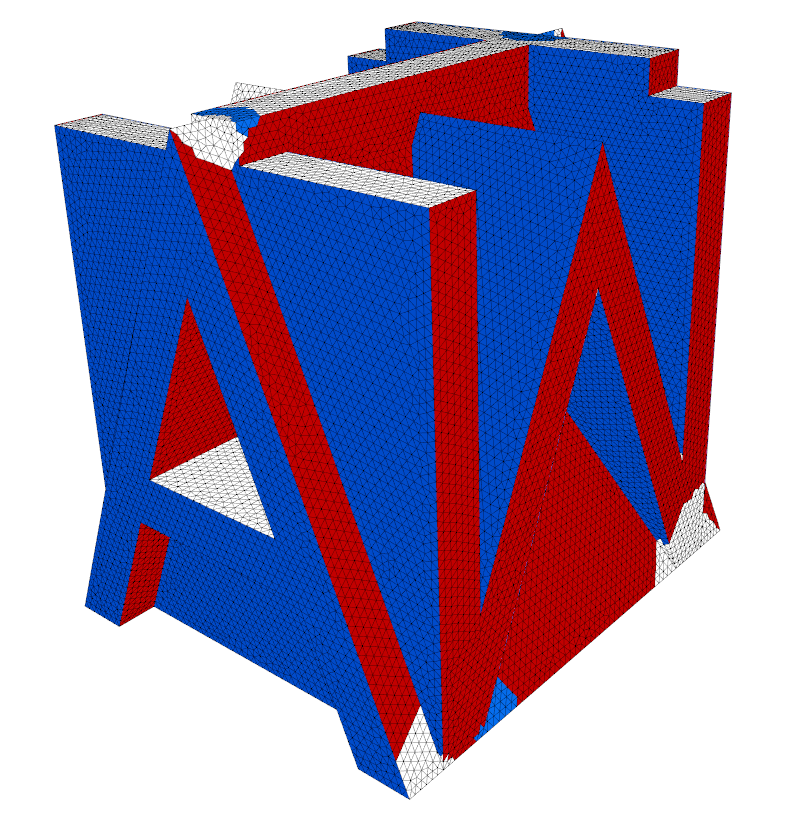}\label{fig_lt2}}\hfill
\subfloat[500 generations]{\includegraphics[width=0.3\hsize]{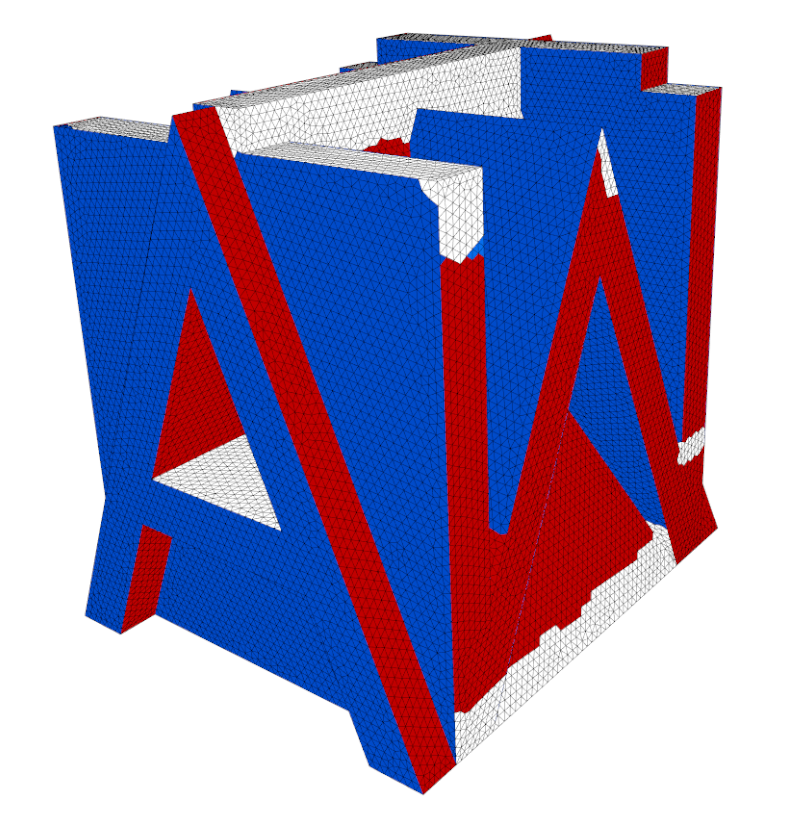}\label{fig_lt3}}
\caption{In complex cases, a large number of generations may not be sufficient to reach a satisfying labeling. On this model (ABC \#4035), sharp wedges are numerous and poorly aligned with base axes. After 500 generations, $V_p(l)$ is reduced from 45 to 26, but many invalidities remain and some of the solutions found are not satisfying.}
\label{fig:longtime}
\end{figure}

The consistency and reproducibility of our results are hindered by the non-deterministic nature of our approach. This drawback can be mitigated by increasing the number of individuals $N$ per generation. More mutations are considered each generation, leading to a greater pool to choose from for the next and an improved breadth of search overall. In contrast, the number of generations $G$ relates to the total number of mutations that can be applied, analogous to the depth of search. In Figure \ref{fig:longtime}, we observe that a very high number of generations does not necessarily imply better results.

In complex cases and similarly to previous work \cite{polycut_livesu}, the final labeling may assign the same label across some feature-edges. In such cases, the feature-edge does not appear in the polycube and is most often deteriorated in the resulting hex-mesh, as illustrated in Figure \ref{fig:bad_feature_edge}. For CAD models, this issue could be alleviated in future work through feature-aware mutations and fitness. 

In our experiments, we further observe that when a vastly superior mutation is found, a single individual and its offspring tend to overwhelm the archive, reducing population diversity. For the most complex shapes, advanced evolutionary mechanisms such as \textit{speciation} \cite{Deb1997PopulationSC} could be added to favor solution diversity.

%% file: p7-conclusion.tex
\section{Conclusion and future work}

We have presented \textit{Evocube}, a novel evolutionary approach to compute polycube labelings. Our method generates pseudo-valid labelings on a wide range of CAD and smooth models, paving the way for future work on labeling enhancements within this framework. 
Resulting polycubes have meaningful complexity and produce high-quality hex meshes on a wide range of input geometries. 


Our method has a few limitations which may be addressed in future work. 
Additional mutations are required to further improve our pseudo-valid labelings. In particular, boundary displacement and feature-aware operations would integrate well in \textit{Evocube} and produce more accurate hex meshes while greatly reducing parameterization distortion. 
Tailor-made mutations may also be added to systematically deal with specific CAD features, such as slopes or sharp wedges.  Finally, we hope our use of an evolutionary algorithm applied to polycube labeling can inspire similar work on other problems in mesh generation and computer graphics in general.


%% file: p8-appendix.tex
\appendix
\section{Validity limitations} \label{app:validity}

\begin{figure}[htb]
\centering
\subfloat[Not necessary]{\includegraphics[width=0.35\hsize]{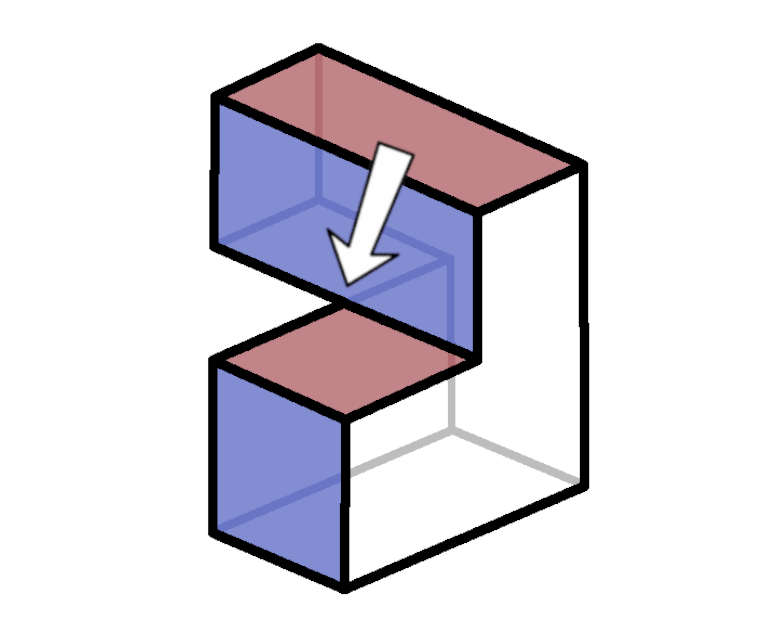}\label{fig_val_fail1}}\hfill
\centering
\subfloat[Not sufficient]{\includegraphics[width=0.35\hsize]{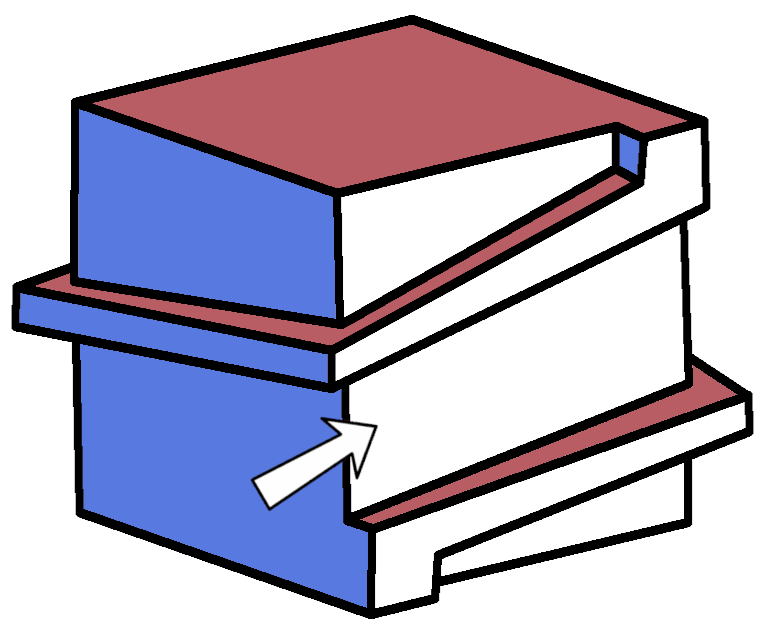}\label{fig_val_fail2}}
\caption{Failure cases of pseudo-validity conditions.}
\label{fig:failure_validity}
\end{figure}

The pseudo-validity conditions adopted by our method are not necessary for the existence of a corresponding polycube polyhedron. A counterexample is shown in Figure \ref{fig:failure_validity}a, where a labeling not fulfilling the conditions still corresponds to a valid polycube polyhedron. The conditions are not sufficient either, as illustrated by the counterexample in Figure \ref{fig:failure_validity}b \cite{sokolov2015fixing}, where a labeling fulfills the conditions but does not correspond to a valid polycube polyhedron.

Our pseudo-validity conditions are a simplification of the ones described by Eppstein and Mumford \cite{eppstein2010steinitz}, adapting the Steinitz criteria \cite{steinitz} for convex polyhedra. Zhao et al. \cite{polycube_shape_space} managed to extend their analysis to orthogonal polyhedra with higher genus. The main advantage of our validity proxy is that it can efficiently be evaluated, allowing for optimization via a heuristic.

\section{Time measurements} \label{app:time}

In Table \ref{tab:detailed_timings}, we report detailed time measurements illustrated in Figure \ref{fig:time_plot}. We measure time both in CPU and real time, and observe a 14 fold speedup. All computations are performed on a 16-core AMD Ryzen Threadripper 1950X 2.2 GHz and 128 Gb RAM.

\begin{table}[htb]
\small
\centering
\begin{tabular}{|l|l|l|}
\hline
Operation & CPU time (h) & Real time (h) \\ \hline
Pre-optimization computation & 0.33 & 0.33 \\ 
Individual selection & 3.20 & 0.21 \\ 
Individual mutations & 7.63 & 0.51 \\ 
Charts and turning points & 6.25 & 0.42 \\ 
Fitness evaluation & 55.2 & 3.66 \\ 
Crossing & $10^{-3}$ & $10^{-3}$ \\ 
Insertion in archive & $10^{-4}$ & $10^{-4}$ \\ 
Post-optimization & 0.21 & 0.21 \\ \hline 
Total & 72.82 & 5.34 \\ 
\hline 

\end{tabular}
\caption{\label{tab:detailed_timings}
         Labeling optimization timings over all 1315 models.}
\end{table}